\documentclass[twocolumn,superscriptaddress,showpacs,notitlepage]{revtex4-1}
\usepackage{amsmath,amssymb}
\usepackage{graphicx}
\usepackage{txfonts}
\usepackage{color}
\usepackage{hyperref}
\usepackage{framed}
\usepackage{wrapfig}
\usepackage[letterpaper]{geometry}
\usepackage{soul}
 \definecolor{ashgrey}{rgb}{0.7, 0.75, 0.71}

\DeclareMathOperator\sech{sech}

\begin{document}

\title{Nonperturbative nonlinear effects in the dispersion relations for TE and TM plasmons on two-dimensional materials}

\author{Vera Andreeva}
\email[Correspondence should be addressed to: ]{andr0616@umn.edu}
\affiliation{Physics Department and International Laser Center, Lomonosov Moscow State University, Moscow 119991, Russia}
\affiliation{School of Mathematics, University of Minnesota, Minneapolis 55455, USA}
\author{Mitchell Luskin}
\affiliation{School of Mathematics, University of Minnesota, Minneapolis 55455, USA}
\author{Dionisios Margetis}
\affiliation{Institute for Physical Science and Technology, Department of Mathematics,
and Center for Scientific Computation and Mathematical Modeling, University of Maryland, College Park, Maryland 20742, USA}

\begin{abstract}
We analytically obtain the dispersion relations for transverse-electric (TE) and transverse-magnetic (TM) surface plasmon-polaritons in a nonlinear two-dimensional (2D) conducting material  with inversion symmetry  lying between two Kerr-type dielectric media.
To this end, we use Maxwell's equations within the quasi-electrostatic, weakly dissipative regime.
We show that the wavelength and propagation distance of surface plasmons decrease due to the nonlinearity of the surrounding dielectric.
In contrast, the effect of the nonlinearity of the 2D material depends on the signs of the real and imaginary parts of the third-order conductivity.
Notably, the dispersion relations obtained by naively replacing the permittivity of the dielectric medium by its nonlinear counterpart in the respective dispersion relations of the linear regime are not accurate.
We apply our analysis to the case of doped graphene and make predictions for the TM-polarized surface plasmon wavelength and propagation distance.
\end{abstract}

\maketitle

\section{Introduction}
\label{sec:Intro}
Surface plasmon-polaritons (SPs) are fine-scale electromagnetic waves bound to the interface between a metal or semimetal and a dielectric~\cite{szunerits2015introduction}.
A striking property of SPs is their possible confinement near atomically thick conducting materials beyond the classical diffraction limit \cite{low2017polaritons,gramotnev2010plasmonics}.
This property has motivated a plethora of exciting applications, giving rise to the active field of plasmonics for two-dimensional (2D) materials~\cite{koppens2011, grigorenko2012, garcia2014, low2014plasmons}.
The high confinement and tunability of SPs has been reported in experiments~\cite{ooi2017,novoselov2004, grigorenko2012}.
This tunability has enabled the fabrication of novel nanophotonic devices~\cite{bao2012, rodrigo2015, low2014}.

Recent experimental developments in using high-power sources in the mid- and far-infrared frequency range~\cite{shalaby2015} pave the way to extensions of plasmonics to the nonlinear regime of the materials involved~\cite{ooi2017}.
A main goal is to utilize nonlinear optical properties of the dielectric substrate and the conducting 2D material in order to increase stability and localization of SPs ~\cite{wang2012, gorbach2013}.
As a result, new, nonlinear SP modes may appear along the 2D material~\cite{bludov2014, wu2017, nesterov2013graphene}. Such modes do not exist in linear media.

In this paper, motivated by the promise of nonlinear plasmonics, we aim to describe the {\em combined effect} of the nonlinearities in both the 2D material and the ambient dielectric media on the dispersion of the SPs.
We separately examine the cases with transverse-electric (TE) and transverse-magnetic (TM) polarization of the SPs by the use of analytical methods.

There are a number of comprehensive theoretical studies that focus on the nonlinear optical response of graphene \cite{mikhailov2007, cheng2014third} as well as black phosphorus~\cite{youngblood2016layer}.
Notably, the magnitude of the nonlinear susceptibility reported for graphene is at least as large as the one of conventional nonlinear materials, such as GaAs ~\cite{mikhailov2007, khurgin2014, cheng2014third, mikhailov2017, khurgin2017}.
Applications of the nonlinear properties of graphene include, but are not limited to, enhancement of third-harmonic generation \cite{nasari2016electrically}, optical bistability \cite{peres2014optical}, solitons \cite{nesterov2013graphene, smirnova2014dissipative} and nonlinear graphene plasmonic waveguides \cite{wang2012, hajian2014, nasari2015, wu2017, qasymeh2017, gorbach2013, bludov2014}.

In this paper, we investigate the compound effect of nonlinearities on the dispersion relation of SPs propagating on isotropic 2D materials  with inversion symmetry.
Our approach recognizes that, in principle, both the 2D conducting material and the surrounding dielectric media may exhibit a nonlinear optical response when irradiated by the (sufficiently strong) electromagnetic field generated by a high-power source.
We invoke time-harmonic Maxwell's equations by restricting attention to the single-frequency response of materials. Hence, phenomena related to frequency generation lie beyond our present scope.

In our analysis, we use a nonperturbative technique for the investigation of the differential equations for the field components. This approach yields the SP dispersion relation analytically in the quasi-electrostatic regime, revealing the exact contribution of the dielectric nonlinearity to the SP (complex) wavenumber.
It should be noted that our result for the SP dispersion relation in terms of conductivity holds only under the assumption of a 2D material with inversion symmetry, e.g., graphene,  even-layered MoS$_2$, black phosphorus.
We discuss in some detail the dispersion of SPs in the particular case of doped graphene by making use of available conductivity models ~\cite{cheng2014third, cheng2015third, mikhailov2016, mikhailov2007} for the nonlinear optical response of this material.

Further, we show that the combined effect of the dielectric and 2D material nonlinearities depends on the signs of the real and imaginary parts of the third-order conductivity of the 2D material.
According to our prediction, the wavelength and propagation distance of SPs may in principle decrease or increase in comparison to the corresponding case of linear media, or even experience no change at all.
In particular, for highly doped graphene in the THz and far-infrared frequency range, the dielectric and graphene nonlinearities cause an increase of the wavelength and propagation distance of the TM-polarized SP.
At the risk of redundancy, we repeat that in this paper we choose not to examine high-harmonic and supercontinuum generation, as well as other nonlinear phenomena related to frequency conversion.

By comparing our present work to recent literature in nonlinear plasmonic systems, we believe that, in a nutshell, other theoretical studies can be separated into two main categories. These focus on {\em either} the dielectric {\em or} the 2D material nonlinearity, but not on both.
Specifically, in studies of the former category, only the dielectric medium surrounding the graphene sheet is assumed to interact in a nonlinear fashion with the light source, while the optical response of the 2D material (usually graphene) is modeled in the linear regime; see, e.g.,~\cite{wang2012, hajian2014, bludov2014, nasari2015, nasari2016, hajian2016guided, wu2017, qasymeh2017, hajian2014}.
In studies of the latter category, only the nonlinearity of the 2D material is examined, while the ambient media are considered as linear; e.g., in~\cite{yao2014, gong2015, mikhailov2017influence}.
In contrast, in our approach the nonlinearities of all materials involved are treated simultaneously.

We should add that dispersion relations for SPs in previous works have been obtained analytically under special assumptions.
One of the most common assumptions for both TE- and TM- polarized SPs has been that of dissipationless
propagation~\cite{hajian2014, hajian2016guided, bludov2014, wu2017, qasymeh2017}.
Another approach involves a perturbation expansion of Maxwell's equations and treats the nonlinearities of the dielectric and 2D material as small~\cite{gorbach2013}.
Our present treatment differs from previous investigations in the following aspects.
First, we systematically consider the case with weak dissipation, thus relaxing the assumptions in~\cite{hajian2014, hajian2016guided, bludov2014, wu2017, qasymeh2017}.
Second, in contrast to~\cite{gorbach2013}, we apply a nonperturbative approach that circumvents the need to treat the nonlinearities as small.

In contrast to the case of TE-polarized SPs, the analytical investigation of TM-polarized SPs is deemed as complicated:
This case is described by a system of coupled differential equations for two electric field components which have a nonzero phase difference.
The simplest scenario of solution arises when the electric field components have a phase difference equal to $\pi/2$~\cite{hajian2014, hajian2016guided}.
In this special case, which we show corresponds to no dissipation, the SP dispersion relation has been derived analytically, since the resulting system of differential equations is integrable~\cite{berkhoer1970self}.
Notably, our analysis transcends this phase limitation.

In this paper, we analytically derive the dispersion relation of TE- and TM-polarized SPs from Maxwell's equations by using a reduced set of assumptions.
First, as we discuss above, we take into account the nonlinearities of the ambient dielectric media {\em and} the 2D conducting material.
Second, we consider small yet nonzero dissipation of the SP propagation; and (only for the case with TM-polarization) apply the quasi-electrostatic approximation, which means that the SP wavenumber is considered as much larger in magnitude than the wavenumbers of the ambient media.
Furthermore, in our approach the effects of the nonlinearities of the dielectric media are not regarded as small and are treated nonperturbatively in the dispersion relation.
This type of treatment allows us to find the dispersion relation of SPs excited by a sufficiently strong electric field.

The remainder of the paper is organized as follows.
In Sec.~\ref{sec:model}, we introduce the geometry along with Maxwell's equations for the problem under study.
In addition, in Sec.~\ref{sec:model} we review the linear case for the convenience of the reader, and for the sake of later comparisons.
Section~\ref{sec:TE} focuses on the dispersion of the TE-polarized SP in the nonlinear regime.
In Sec.~\ref{sec:TM}, we address the more demanding problem of the corresponding dispersion relation for the TM-polarized SP.
Section~\ref{sec:discussion} contains a discussion of our predictions for the particular system of doped graphene. Section~\ref{sec:conclusion} concludes the paper with a summary of the main results and an outline of open problems.
The appendices provide technical derivations needed in the main text.
Throughout this paper, we assume that the fields have the temporal dependence $e^{-i\omega t}$, where $\omega$ is the radial frequency.
We use the centimetre-gram-second (CGS) system of units.

\section{Model and geometry}
\label{sec:model}

In this section, we describe the geometry and governing equations of the problem under consideration.
By focusing on the single-frequency response of materials, we use the time-harmonic Maxwell equations along with suitable (transmission) boundary conditions for the electromagnetic field on the 2D material sheet.

\begin{figure}[h]
\center{\includegraphics[width=1\linewidth]{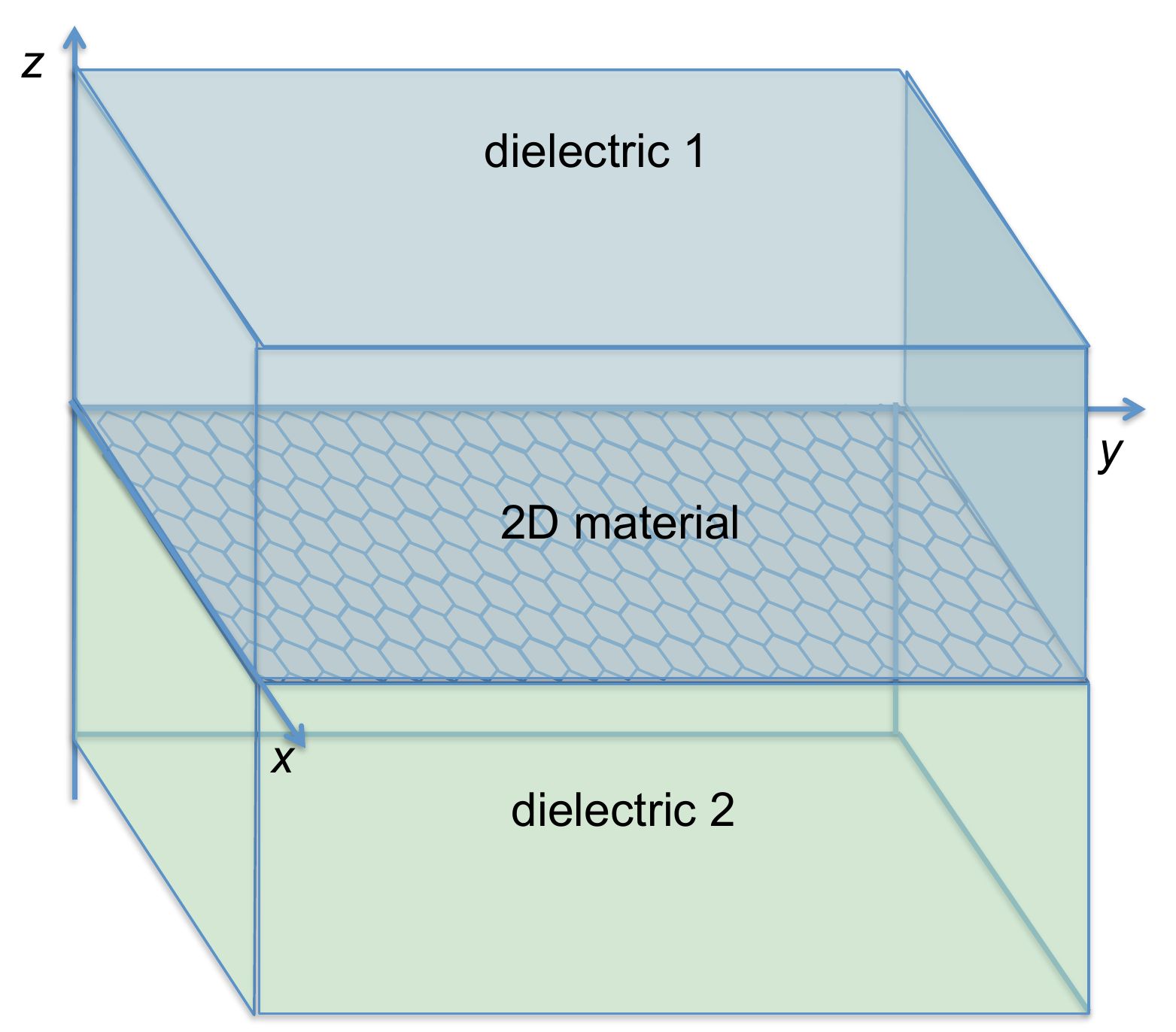}}
\centering{}\caption{(Color online) Geometry of the system under investigation.
The flat  material sheet lies at the interface (plane at $z=0$) between two unbounded dielectric media.}
\label{fig:sk}
\end{figure}

In our setting, the conducting sheet lies on the $xy$-plane, between two unbounded dielectric media, as shown in Fig. \ref{fig:sk}.
We choose the positive $x$-axis as the direction of the SP propagation.
The ambient medium $j$ has dielectric permittivity relative to the vacuum equal to $\epsilon_j$, where $j=1$ for the upper half space, $z>0$, and $j=2$ for the lower half space, $z<0$.

In the absence of external current-carrying sources, the curl laws of Maxwell's equations in the dielectric media are given by
\begin{align}
\label{m1}
\nabla\times\bold{H}_{j}&=-\frac{i\omega}{c}\bold{D}_{j},\\
\label{m2}
\nabla\times\bold{E}_{j}&=\frac{i\omega}{c}\bold{H}_{j}\qquad (j=1,\,2).
\end{align}
In the above, $\bold{E}_{j}\left(x,y,z\right)$, $\bold{H}_{j}\left(x,y,z\right)$ and $\bold{D}_{j}\left(x,y,z\right)$ are the electric, magnetic and displacement fields, respectively, and $c$ is the speed of light in vacuum.
Here, we assume that the ambient media are non-magnetic.

Equations~\eqref{m1} and~\eqref{m2} should be supplemented with the suitable (radiation) condition at large distance from the material sheet as $|z|\rightarrow\infty$.
Since we single out the SP as an evanescent wave, we require that the electromagnetic field should decay as $|z|\to\infty$.
In addition, at the planar interface ($z=0$) we impose: (i) the continuity of the tangential component of the electric field; and (ii) a jump condition in the tangential component of the magnetic field that accounts for the surface current, $\bold{j}_{s}$, induced by the tangential electric field on the sheet.
These conditions explicitly are
\begin{gather}
\label{bc1}
\left(\bold{H}_1-\bold{H}_2\right)\times\bold{n}=\frac{4\pi}{c}\bold{j}_{s},\\
\label{bc2}
\left(\bold{E}_1-\bold{E}_2\right)\times\bold{n}=\bold{0},
\end{gather}
where $\bold{n}=-\bold{e}_z$ is the ($z$-directed) unit vector perpendicular to the sheet that points downwards.
For our purposes, $\bold{j}_s$ is in principle a functional of $\bold{E}_{j}(x,y,0)\times \bold{n}$ which is single-valued on the sheet.
For details, we refer the reader to Secs.~\ref{subsec:plasmons} and~\ref{subsec:nonl-SP}.

\subsection{Revisiting SPs in the linear regime}
\label{subsec:plasmons}

Next, we review the dispersion relations for TE- and TM-polarized SPs in the case with a linear conducting sheet and linear ambient dielectrics.
For sufficiently small magnitude of the electric field, $\bold{E}_{j}$, the relation between the displacement field, $\bold{D}_j$, and $\bold{E}_j$ can be approximated by \cite{shen1984}
\begin{equation}
\label{cr1}
\bold{D}_j=\epsilon_j\bold{E}_j,
\end{equation}
where $\epsilon_j$ is the constant dielectric permittivity of medium $j$ ($j=1,\, 2$).
In this vein, the surface current, $\bold{j}_{s}$, induced on the conducting sheet obeys the linear relation
\begin{equation}
\label{sc1}
\bold{j}_{s}=\sigma^{(1)}\bold{E}_\parallel.
\end{equation}
In the above, $\bold{E}_\parallel=\bold{e}_x E_x +\bold{e}_y E_y$ is the electric field tangential to the sheet at $z=0$, $\sigma^{(1)}\equiv \sigma^{(1)}_{xx}=\sigma^{(1)}_{yy}$ is the first-order surface conductivity of the 2D material~\cite{falkovsky2007, mikhailov2016}, and $\bold{e}_\ell$ denotes the $\ell$-directed unit Cartesian vector ($\ell=x,\,y$).
Here, we consider an isotropic sheet; thus, $\sigma^{(1)}=\sigma^{(1)}(\omega)$ is a scalar function of frequency, $\omega$.
To account for energy dissipation in the 2D material, we need to have ${\rm Re} \,\sigma^{(1)}(\omega)>0$.

In this framework, the dispersion relation for SPs can be found via particular solutions of Eqs.~\eqref{m1}--\eqref{bc2} that behave as $e^{ik_x x}$ in $x$ by using relation~\eqref{sc1}.
The associated wavenumber, $k_x$, is determined as a function of frequency, $\omega$.
For a TE-polarized SP, the only nonzero components of the electromagnetic field are $H_x$, $H_z,$ and $E_y$; whereas a TM-polarized SP corresponds to nonzero $E_x$, $E_z,$ and $H_y$.

In particular, the dispersion relation for the TE-polarized SP including retardation is~\cite{bludov2013primer}
\begin{equation}
\label{eq:TE-lin-disp}
\sqrt{k_x^2-\frac{\omega^2\epsilon_1}{c^2}}+\sqrt{k_x^2-\frac{\omega^2\epsilon_2}{c^2}}=\frac{4\pi\omega i\sigma^{(1)}}{c^2}.
\end{equation}
This equation is subject to the radiation condition which in turn implies the constraint ${\rm Re}\sqrt{k_x^2-(\omega/c)^2\epsilon_j}>0$ ($j=1,\,2$)~\cite{bludov2013primer}.
Evidently, for lossless surrounding media, i.e., positive $\epsilon_j$, Eq.~\eqref{eq:TE-lin-disp} has an admissible solution for $k_x$ if ${\rm Im}\,\sigma^{(1)}<0$.

On the other hand, the dispersion relation for the TM-polarized SP is given by~\cite{bludov2013primer}
\begin{equation}
\label{eq:TM-lin-disp}
\frac{\epsilon_1}{\sqrt{k_x^2-\omega^2\epsilon_1/c^2}}+\frac{\epsilon_2}{\sqrt{k_x^2-\omega^2\epsilon_2/c^2}}=\frac{4\pi\left(-i\sigma^{(1)}\right)}{\omega}.
\end{equation}
Because we impose ${\rm Re}\,\sqrt{k_x^2-(\omega/c)^2\epsilon_j}>0$, Eq.~\eqref{eq:TM-lin-disp} has a solution for $k_x$ if ${\rm Im}\,\sigma^{(1)}>0$ in regard to lossless dielectrics.
Hence, in view of the mutually incompatible restrictions on $\sigma^{(1)}$, one sees that it is impossible to excite at a given frequency both a TE- and a TM-polarized SP on a linear 2D material lying between two lossless media.

It is of physical interest to discuss the dispersion relation for the TM case in the quasi-electrostatic regime, when the wavenumber of the SP is much larger in magnitude than the wavenumber in free space, viz., $|k_x|>>\omega/c$.
This possibility is afforded by Eq.~\eqref{eq:TM-lin-disp} if
$|\sigma^{(1)}|$ is sufficiently small with ${\rm Im}\,\sigma^{(1)}>0$.
Accordingly, under the assumption that
$c(\epsilon_1+\epsilon_2)/|\sigma^{(1)}|\gg 4\pi $, Eq.~\eqref{eq:TM-lin-disp} yields~\cite{bludov2013primer}
\begin{equation}
\label{eq:TM-quasi}
k_x\approx\frac{\omega\left(\epsilon_1+\epsilon_2\right)}{4\pi\left(-i\sigma^{(1)}\right)}.
\end{equation}
By this formula, $k_x^r\equiv {\rm Re}\,k_x >0$ and $k_x^i\equiv {\rm Im}\,k_x> 0$; thus, the TM-polarized SP propagates and decays (for a dissipative sheet) in the positive $x$-direction.

A figure of merit for the TM-polarized SP is the ratio $k_x^i/k_x^r$, which expresses the (relative) damping of this wave in the direction of propagation~\cite{low2014plasmons}.
By inspection of Eq.~\eqref{eq:TM-quasi}, we find that $k_x^i/k_x^r=\sigma_r^{(1)}/\sigma_i^{(1)}$.
Thus, it is desirable to use frequencies at which $\left|\sigma_r^{(1)}/\sigma_i^{(1)}\right|\ll 1$.
This condition defines the weakly dissipative regime in the linear case.

We now turn our attention to the TE-polarized SP. By Eq.~\eqref{eq:TE-lin-disp} with ${\rm Im}\,\sigma^{(1)}<0$, the related wavenumber is~\cite{bludov2013primer}
\begin{equation}
\label{eq:TE-disp-ex}
k_x=\frac{\omega}{c}\left(\frac{\epsilon_1+\epsilon_2}{2}-
\frac{4\pi^2\left(\sigma^{(1)}\right)^2}{c^2}-\frac{c^2\left(\epsilon_1-\epsilon_2\right)^2}
{64\pi^2\left(\sigma^{(1)}\right)^2}\right)^{1/2}.
\end{equation}
By Eq.~\eqref{eq:TE-disp-ex}, $|k_x|$ may become much larger than $\omega/c$ if $2\pi|\sigma^{(1)}|/c\gg 1$, assuming that $\epsilon_j$ is close to unity for each $j$.
In contrast, one obtains $k_x\approx (\omega/c)\sqrt{(\epsilon_1+\epsilon_2)/2}$ provided
\begin{equation*}
	\frac{\epsilon_1+\epsilon_2}{2} \gg \max\left\{\frac{4\pi^2}{c^2}|\sigma^{(1)}|^2,\frac{c^2\left(\epsilon_1-\epsilon_2\right)^2}
{64\pi^2|\sigma^{(1)}|^2}\right\}.
\end{equation*}

The SP wavenumbers from the above dispersion relations can be manipulated via the tuning of $\sigma^{(1)}$.
For example, in the case of highly doped graphene, the Fermi energy, $E_F$, is much larger than the Boltzmann energy, $k_BT$.
Accordingly, the surface conductivity, $\sigma^{(1)}(\omega)$, at the THz and far-infrared frequency ranges has the Drude form~\cite{falkovsky2007, mikhailov2016}
\begin{gather}
\label{eq:Drude}
\sigma^{(1)}(\omega)=\frac{i\sigma_0^{(1)}}{\Omega+i\Gamma},
\end{gather}
where $\Omega=\hbar\omega/E_F$ and $\Gamma=\hbar\gamma/E_F$ are non-dimensional parameters, $\sigma_0^{(1)}=e^2/(\hbar\pi)$ has units of surface conductivity, $e$ is the electron charge, and $\hbar$ is Planck's constant.
In addition, $\gamma$ is the phenomenological relaxation rate due to the scattering of electrons by impurities, phonons, and lattice imperfections~\cite{mikhailov2016}.
By changing the doping of graphene, one can control $\sigma^{(1)}$.
Therefore, by Eqs.~\eqref{eq:TM-quasi}--\eqref{eq:Drude} the SP wavenumber, $k_x$, can be manipulated through doping~\cite{novoselov2004, grigorenko2012}.

In graphene, energy losses due to the scattering of electrons by other particles can be considered as relatively low; thus, $\Gamma\ll 1$~\cite{mikhailov2016}.
Furthermore, it is possible to have ${\rm Im}\,\sigma^{(1)}>0$ at a suitable frequency range, which in turn allows the propagation of the TM-polarized SP.
This SP can exhibit a weak decay in doped graphene at low enough frequencies, in a regime where Eq.~\eqref{eq:Drude} presumably holds.
Recall that the TM- and TE-polarized SP may not be simultaneously present in graphene.
For higher frequencies, the TE-polarized SP can exist in a narrow frequency range depending on the optical contrast, $\epsilon_1-\epsilon_2$, of the surrounding dielectric media~\cite{kotov2013ultrahigh}.

\subsection{Model in the nonlinear regime}
\label{subsec:nonl-SP}

Next, we address the possible appearance of SPs by taking into account nonlinearities in both the 2D material and the ambient dielectrics.
We recognize that when a sufficiently strong electric field, $\bold{E}_j$, is present, the response of the corresponding media may not be described by linear constitutive law~\eqref{cr1} and surface current~\eqref{sc1}.
Instead, one must invoke the nonlinear constitutive law between $\bold{D}_j$ and $\bold{E}_j$, in combination with a nonlinear relation $\bold{j}_s$ and $\bold{E}_j$.

To describe this nonlinear response, we assume that the dielectric media are isotropic and centrosymmetric.
Accordingly, their second-order nonlinear response vanishes; and the constitutive relation that describes the third-order Kerr-type nonlinearity is given by \cite{shen1984}
\begin{equation}
\label{eq:D-nonl}
\bold{D}_{j}=\epsilon_j\bold{E}_{j}+4\pi\chi^{(3)}_j\left|\bold{E}_{j}\right|^2\bold{E}_{j},
\end{equation}
where $\chi^{(3)}_j$ is the third-order susceptibility of dielectric $j$.
Throughout this paper, we assume that $\chi^{(3)}_j>0$ $(j=1,2)$.

In a similar vein, we consider the nonlinear response of the 2D material.
By considering an isotropic conducting sheet  with inversion symmetry, we invoke the following relation for the surface current:
\begin{equation}\label{eq:js-nonl}
\bold{j}_{s}=\left(\sigma^{(1)}+\sigma^{(3)}\left|\bold{E}_\parallel\right|^2\right)\bold{E}_\parallel,
\end{equation}
where $\sigma^{(3)}\equiv\sigma^{(3)}_{xxxx}=\sigma^{(3)}_{yyyy}$ is the third-order conductivity of the conducting sheet~\cite{mikhailov2007, cheng2015third, mikhailov2016}.
Recall that $\bold{E}_\parallel$ is the electric field tangential to the sheet at $z=0$.

In Eq.~\eqref{eq:js-nonl}, the parameters $\sigma^{(l)}$ ($l=1,\,3$) are of course frequency ($\omega$-) dependent. In Secs.~\ref{sec:TE} and~\ref{sec:TM}, we derive $k_x$ as a function of these parameters, $\sigma^{(1)}$ and $\sigma^{(3)}$. This result is general within a class of 2D materials, i.e., the materials with inversion symmetry. In more detail, we obtain a dispersion relation, describing SPs in 2D materials for which the following assumptions hold: (1) the second-order nonlinear conductivity is negligible, and (2) the real part of the effective conductivity, $\sigma^{(1)}+\sigma^{(3)}\left|\bold{E}_\parallel\right|^2,$ is small compared to the imaginary part. Examples of such materials are graphene, black phosphorus, and even-layered MoS$_2$~\cite{li2013probing}. It should be noted that Eq.~\eqref{eq:js-nonl} does not describe 2D materials with broken inversion symmetry, such as odd-layered transition metal dichalcogenides, which can exhibit a strong second-harmonic generation~\cite{li2013probing}.

In particular, the third-order conductivity, $\sigma^{(3)}$, of  doped graphene  in the THz and far-infrared frequency ranges has been obtained via a quantum-mechanical approach~\cite{cheng2014third, cheng2015third, mikhailov2016} and a kinetic treatment based on the Boltzmann equation~\cite{mikhailov2007}.
This parameter is expressed by the formula
\begin{equation}
\label{eq:Drude3}
\sigma^{(3)}(\omega)=-\frac{i\sigma^{(3)}_0}
{\left(\Omega+i\Gamma\right)^2\left(\Omega-i\Gamma\right)},
\end{equation}
where $\sigma^{(3)}_0=e^4\hbar v_F^2/(8\pi E_F^4)$, $v_F\approx 10^8$ cm/s is the Fermi velocity, and $\Omega$ and $\Gamma$ are defined in the context of Eq.~\eqref{eq:Drude}.
Equation~\eqref{eq:Drude3} describes the nonlinear response of graphene at the frequency, $\omega$, of the
incident wave.
In general, $\sigma^{(3)}$ is a function of three distinct frequencies, and is responsible for frequency mixing processes~\cite{mikhailov2016} which are beyond the scope of this work.
Similar to the linear case (Sec.~\ref{subsec:plasmons}), in graphene $\sigma^{(3)}$ can be controlled
via doping~\cite{cheng2014third}.
 Note that Eq.~\eqref{eq:Drude3} is based on the assumption that the carbon atoms are arranged in a honeycomb lattice ~\cite{cheng2014third, cheng2015third, mikhailov2016} and the energy spectrum of the 2D electron/hole gas is linear~\cite{mikhailov2007}.

A remark on possible approximations associated to Eq.~\eqref{eq:Drude3} is in order.
Define $\sigma_r^{(l)}$ ($\sigma_i^{(l)}$) as the real (imaginary) part of $\sigma^{(l)}$ for $l=1,\,3$.
In the weakly dissipative regime considered here, the real part of the total conductivity, $\sigma_r=\sigma_r^{(1)}+\sigma_r^{(3)}|\bold{E}_\parallel|^2$, which expresses the losses in the 2D material, is small compared to the respective imaginary part, $\sigma_i=\sigma_i^{(1)}+\sigma_i^{(3)}|\bold{E}_\parallel|^2$.
Hence, one may apply the condition $\left|\sigma_r/\sigma_i\right|\ll 1$ in the appropriate frequency range.
For graphene, this assumption holds when $\hbar\omega<2E_F$ and the doping is high, which implies $E_F\gg k_BT$  ~\cite{mikhailov2016}.

It is worthwhile to entertain the following naive scenario of obtaining the dispersion relations for SPs in the nonlinear regime: Suppose that one simply replaces the dielectric permittivity $\epsilon_j$ by its modified, nonlinear version $\epsilon_j+4\pi\chi^{(3)}_j\left|\bold{E}\right|^2$ in Eqs.~\eqref{eq:TM-quasi} and~\eqref{eq:TE-disp-ex}; and analogously for $\sigma^{(1)}$.
We will show that this approach provides incorrect results both for the TE- and TM-polarized SPs (Secs.~\ref{sec:TE} and~\ref{sec:TM}).

\section{TE-polarized surface plasmon}
\label{sec:TE}

In this section, we derive the dispersion relation of the TE-polarized SP by using the nonlinear model of Sec.~\ref{subsec:nonl-SP}.
For this purpose, we apply approximations subject to the assumption of weak dissipation, according to which the imaginary part, $k_x^i$, and real part, $k_x^r$, of the SP wavenumber, $k_x$, satisfy $0<k_x^i\ll k_x^r$.
We remind the reader that we use the convention of wave propagation along the positive $x$-axis, thus taking $k_x^r$ and $k_x^i$ to be positive.

In the present case with TE-polarization, the electric, displacement, and magnetic fields are
\begin{align*}
\bold{E}_j\left(x,z\right)&=\left(0, E_{jy}\left(z\right), 0\right)e^{ik_xx},\\
\bold{D}_j\left(x,z\right)&=\left(0, D_{jy}\left(z\right), 0\right)e^{ik_xx},
\\
\bold{H}_j\left(x,z\right)&=\left(H_{jx}\left(z\right), 0, H_{jz}\left(z\right)\right)e^{ik_xx}\quad (j=1,\,2),
\end{align*}
where $z>0$ for $j=1$ and $z<0$ for $j=2$. Notice that $\bold{E}_\parallel=\bold{e}_y E_{1y}=\bold{e}_y E_{2y}$ at $z=0$ in this setting; cf. Eq.~\eqref{bc2}.
Substituting the above expressions for $\bold{E}_j$, $\bold{D}_j,$ and $\bold{H}_j$ into Eqs.~\eqref{m1} and~\eqref{m2}, we obtain the following system of equations for the respective field components:
\begin{subequations}\label{eq:system}
\begin{gather}
\label{m1m11}
\frac{dH_{jx}}{dz}-ik_xH_{jz}=-\frac{i\omega}{c}D_{jy},\\
\label{m3m33}
-\frac{dE_{jy}}{dz}=\frac{i\omega}{c}H_{jx},\\
\label{m2m22}
ik_xE_{jy}=\frac{i\omega}{c}H_{jz}.
\end{gather}
\end{subequations}
By making use of constitutive law~\eqref{eq:D-nonl} for $D_{jy}$ and  eliminating the magnetic field components, we obtain an ordinary differential equation for $E_{yj}$, viz.,
\begin{equation}
\label{eq:ode-Ey}
\frac{d^2E_{jy}}{dz^2}-k_x^2E_{jy}=-k_j^2E_{jy}-k_j^2\varepsilon_j\left|E_{jy}\right|^2E_{jy},\\
\end{equation}
where $k^2_j=\omega^2\epsilon_j/c^2$ and $\varepsilon_j=4\pi\chi^{(3)}_j/\epsilon_j$.
Note that the solution to Eq.~\eqref{eq:ode-Ey} in the non-dissipative regime (when $k_x^i=0$) is obtained in~\cite{shen1984};
and the resulting dispersion relation, $k_x=k_x(\omega)$, is discussed in detail in~\cite{bludov2014, wu2017, qasymeh2017}.

In this work, we aim to extend previous analyses by deriving the SP dispersion relation in the presence of sufficiently small dissipation.
We proceed to simplify Eq.~\eqref{eq:ode-Ey} accordingly.
By writing $E_{jy}=|E_{jy}|e^{i\phi_j}$, we obtain the following  equations for the magnitude, $|E_{jy}|$, and phase, $\phi_j$, of the electric field in dielectric medium $j$:
\begin{subequations}\label{eq:Ey-mag-phase}
\begin{gather}
\label{eq:Ey-mag}
\frac{d^2\left|E_{jy}\right|}{dz^2}-\left|E_{jy}\right|\left(\frac{d\phi_j}{dz}\right)^2=\left(\left(k_x^r\right)^2-k_j^2-k_j^2\varepsilon_j\left|E_{jy}\right|^2\right)\left|E_{jy}\right|,\\
\label{eq:Ey-phase}
2\frac{d\left|E_{jy}\right|}{dz}\frac{d\phi_j}{dz}+\left|E_{jy}\right|\frac{d^2\phi_j}{dz^2}=2k_x^rk_x^i\left|E_{jy}\right|.
\end{gather}
\end{subequations}

To make further progress in simplifying the governing equations, we apply the weak-dissipation expansions
\begin{equation*}
	|E_{jy}|\approx |E_{jy}|^{(0)}+(k_x^i/k_x^r)\, |E_{jy}|^{(1)}\ \mbox{and}\ \phi_j\approx \phi_j^{(0)}+(k_x^i/k_x^r)\, \phi_j^{(1)},
\end{equation*}
which are expected to be meaningful if $|k_x^i/k_x^r|\ll 1$.
In the above, the superscripts of $|E_{jy}|$ and $\phi_j$ denote perturbation order (not to be confused with the superscripts in $\sigma^{(1)}$, $\sigma^{(3)}$ and $\chi_j^{(3)}$).
In particular, $|E_{jy}|^{(0)}=|E_{jy}^{(0)}|$ and $\phi_j^{(0)}$ are the zeroth-order variables for the magnitude and phase of the electric field component, which pertain to the non-dissipative system; while $|E_{jy}|^{(1)}$ and $\phi_j^{(1)}$ denote the first-order counterparts which account for dissipation to leading order in $k_x^i/k_x^r$.
Thus, we assume that $|E_{jy}|^{(1)}$ and $\phi_j^{(1)}$ do not depend on $k_x^i/k_x^r$ as this parameter approaches zero.
By substitution of the weak-dissipation expansions into Eqs.~\eqref{eq:Ey-mag-phase} and application of dominant balance in the parameter $k_x^i/k_x^r$, we obtain two sets of equations, one set for each perturbation order.
Using this result, we obtain
\begin{subequations}\label{eq:dEydz}
\begin{gather}
\label{eq:dEydz-med1}
\begin{split}
&\left.\frac{dE_{1y}}{dz}\right|_{z=0}=-\sqrt{\left(k_x^r\right)^2-k_1^2-\frac{1}{2}k_1^2\varepsilon_1\left|E_0\right|^2}E_0\\
&+\frac{4ik_x^rk_x^iE_0}{k_1^2\varepsilon_1\left|E_0\right|^2}\left(\sqrt{\left(k_x^r\right)^2-k_1^2-\frac{1}{2}k_1^2\varepsilon_1\left|E_0\right|^2}-\sqrt{\left(k_x^r\right)^2-k_1^2}\right),
\end{split}\\
\label{eq:dEydz-med2}
\begin{split}
&\left.\frac{dE_{2y}}{dz}\right|_{z=0}=\sqrt{\left(k_x^r\right)^2-k_2^2-\frac{1}{2}k_2^2\varepsilon_2\left|E_0\right|^2}E_0\\
&-\frac{4ik_x^rk_x^iE_0}{k_2^2\varepsilon_2\left|E_0\right|^2}\left(\sqrt{\left(k_x^r\right)^2-k_2^2-\frac{1}{2}k_2^2\varepsilon_2\left|E_0\right|^2}-\sqrt{\left(k_x^r\right)^2-k_2^2}\right),
\end{split}
\end{gather}
\end{subequations}
where $E_0=\bold{e}_y\cdot \bold{E}_\parallel=E_{1y}(y,0)=E_{2y}(y,0)$ is the value of the electric field on the conducting sheet.
Recall that because of the continuity of the tangential electric field across the sheet, condition~\eqref{bc2}, this $E_0$ is uniquely defined at $z=0$.
For details on the derivation of Eqs.~\eqref{eq:dEydz}, see Appendix~\ref{app:AX}.

By use of boundary conditions~\eqref{bc1} and~\eqref{bc2} along with constitutive law~\eqref{eq:js-nonl} and Eq.~\eqref{m3m33}, we obtain
\begin{equation*}
\left(\frac{dE_{2y}}{dz}-\frac{dE_{1y}}{dz}\right)\Biggl|_{z=0}=\frac{4\pi i\omega}{c^2}\sigma(|E_0|) E_0,
\end{equation*}
where $\sigma(u)= \sigma^{(1)}+\sigma^{(3)}u^2$.
Hence, the {\em normal derivative} of the tangential electric field on the sheet has a {\em jump} proportional to the magnitude of the surface current.
The substitution of the normal derivative of $E_{jy}$ ($j=1,\,2$) in the above jump at $z=0$ by the respective formula of Eqs.~\eqref{eq:dEydz} yields
\begin{subequations}\label{eq:kx-TE}
\begin{widetext}
\begin{gather}
\label{kxr}
k_x^r\approx \frac{\omega}{c}\left(\frac{\epsilon_1+2\pi\chi^{(3)}_1\left|E_0\right|^2+\epsilon_2+2\pi\chi^{(3)}_2\left|E_0\right|^2}{2}+\frac{4\pi^2\sigma_i^2}{c^2}+\frac{c^2\left(\epsilon_1+2\pi\chi^{(3)}_1\left|E_0\right|^2-\epsilon_2-2\pi\chi^{(3)}_2\left|E_0\right|^2\right)^2}{64\pi^2\sigma_i^2}\right)^{1/2}, \\
\label{kxi}
k_x^i\approx \frac{4\pi^2\omega\sigma_r}{c^2\left(k_x^rc/\omega\right)}\left[\frac{\sqrt{\left(k_x^rc/\omega\right)^2-\epsilon_1}-\sqrt{\left(k_x^rc/\omega\right)^2-\epsilon_1-2\pi\chi^{(3)}_1\left|E_0\right|^2}}{\chi^{(3)}_1\left|E_0\right|^2}+\frac{\sqrt{\left(k_x^rc/\omega\right)^2-\epsilon_2}-\sqrt{\left(k_x^rc/\omega\right)^2-\epsilon_2-2\pi\chi^{(3)}_2\left|E_0\right|^2}}{\chi^{(3)}_2\left|E_0\right|^2}\right]^{-1},
\end{gather}
\end{widetext}
\end{subequations}
where $\sigma_r$ ($\sigma_i$) is the real (imaginary) part of $\sigma(|E_0|)$.
In the weakly dissipative regime considered here, the real part, $k_x^r$, of $k_x$ depends only on $\sigma_i$ to leading order in $\sigma_r/\sigma_i$.
In fact, the next-order term for $k_x^r$ is quadratic in $\sigma_r/\sigma_i$.
On the other hand, the imaginary part, $k_x^i$, of $k_x$ is linear in $\sigma_r/\sigma_i$ to leading order in perturbation theory.

For the derivation of formula~\eqref{kxr} we assume that $\sigma_i<0$.
Recall that in the linear regime (in which $\sigma_i^{(3)}=0$) the condition for the appearance of the TE-polarized SP is $\sigma_i^{(1)}<0$ (Sec.~\ref{subsec:plasmons}).

 It is worthwhile to compare Eqs.~\eqref{eq:kx-TE} with corresponding dispersion relations reported in the literature. For example, in \cite{bludov2014} non-dissipative SPs in linear graphene lying between nonlinear Kerr-type and linear dielectric media are studied. It was shown that in this regime a new type of nonlinear surface mode can exist, which does not have a linear counterpart. The dispersion expressed by Eqs.~\eqref{eq:kx-TE} is in agreement with the corresponding relation Eq. (10) in \cite{bludov2014}. In fact, Eq. (10) of \cite{bludov2014} can be obtained from Eq.~\eqref{kxr} by substituting $\sigma_i=\sigma_i^{(1)}$, $\sigma_r=0$ and $\chi^{(3)}_2=0$.

By comparing Eqs.~\eqref{eq:kx-TE} to their linear counterpart, dispersion relation~\eqref{eq:TE-disp-ex}, we make the following observation.
The joint effect of the nonlinearities of the materials on the SP dispersion cannot be accurately captured by simply replacing the dielectric permittivity $\epsilon_j$ in~\eqref{eq:TE-disp-ex} by $\epsilon_j+4\pi\chi^{(3)}_j\left|E_0\right|^2$.
The failure of this naive approach is recognized as follows.
The propagation constant $k_x^r$ includes the effect of the third-order susceptibility, $\chi^{(3)}_j$, with a coefficient equal to $2\pi\left|E_0\right|^2$ instead of the naively expected $4\pi\left|E_0\right|^2$.
Note that for a higher-order nonlinearity the above numerical factor would be different \cite{note} .

For the sake of simplicity, let us assume that the ambient media have the same dielectric properties, viz., $\epsilon_1=\epsilon_2=\epsilon$, $\chi^{(3)}_1=\chi^{(3)}_2=\chi^{(3)}$. According to Eqs.~\eqref{eq:kx-TE}, the damping of the TE-polarized SP can be expressed by the ratio
\begin{equation}
\label{eq:dmp-full}
\frac{k_x^i}{k_x^r}=\frac{\pi\sigma_r}{c\sqrt{\epsilon}}
\frac{\sqrt{2\pi\chi^{(3)}\left|E_0\right|^2/\epsilon+4\pi^2\sigma_i^2/(c^2\epsilon)}+2\pi\left|\sigma_i\right|/(c\sqrt{\epsilon})}{1+2\pi\chi^{(3)}\left|E_0\right|^2/\epsilon+4\pi^2\sigma_i^2/(c^2\epsilon)}.
\end{equation}
Notably, two nonlinear parameters of the effective conductivity of the 2D material, $\sigma_i=\sigma^{(1)}_i+\sigma^{(3)}_i\left|E_0\right|^2$ and $\sigma_r=\sigma^{(1)}_r+\sigma^{(3)}_r\left|E_0\right|^2$, and the nonlinearity of the ambient dielectric, $\chi^{(3)}$, affect the ratio $k_x^i/k_x^r$. In Sec.~\ref{sec:discussion}, we discuss the effect of the dielectric and 2D material nonlinearities on the damping of TE modes in comparison with TM modes for the particular case of doped graphene.

By Eq. \eqref{eq:dmp-full}, $k_x^i/k_x^r$ depends on the ratio $4\pi^2\sigma_i^2/(c^2\epsilon)$. In the quasi-electrostatic regime for the TE mode, $4\pi^2\sigma_i^2\gg c^2\epsilon$, Eq. \eqref{eq:dmp-full} can be simplified to
\begin{equation}
\label{eq:dmp}
\frac{k_x^{i}}{k_x^{r}}\approx\frac{\sigma_r}{\left|\sigma_i\right|}.
\end{equation}
In Sec.~\ref{sec:TM}, we compare Eq.~\eqref{eq:dmp} with the corresponding relation for TM plasmons.

It is of interest to compare the dispersion relation expressed by Eqs.~\eqref{eq:kx-TE} to the corresponding relation in the linear regime, Eq.~\eqref{eq:TE-disp-ex}.
For weak nonlinearities, if $4\pi\chi^{(3)}|E_0|^2/\epsilon_1\ll 1$ and $|\sigma^{(3)}|\,|E_0|^2/|\sigma^{(1)}|\ll 1,$ the real and imaginary parts of the wavenumber of the TE-polarized SP are approximated by
\begin{subequations}\label{eq:kx-TE-same-med}
\begin{equation}
\begin{split}
\label{apr1}
k_x^r\approx\frac{\omega}{c}\left[\epsilon+\frac{4\pi^2\left(\sigma^{(1)}_i\right)^2}{c^2}+\left(\frac{8\pi^2\sigma^{(1)}_i\sigma^{(3)}_i}{c^2}+2\pi\chi^{(3)}\right)\left|E_0\right|^2\right]^{1/2}\\
\approx k_x^{r,{\rm lin}}\left[1+\frac{\omega^2}{2c^2\left(k_x^{r,lin}\right)^2}\left(\frac{8\pi^2\sigma^{(1)}_i\sigma^{(3)}_i}{c^2}+2\pi\chi^{(3)}\right)\left|E_0\right|^2\right],
\end{split}
\end{equation}
\begin{equation}
\label{apr2}
\begin{split}
&k_x^i\approx\frac{4\pi^2\omega\sigma_r^{(1)}\left(-\sigma_i^{(1)}\right)}{c^3\sqrt{\epsilon+\frac{4\pi^2\left(\sigma^{(1)}_i\right)^2}{c^2}}}\\
&\times\left[1+\left(\frac{c^2\chi^{(3)}}{8\pi\left(\sigma_i^{(1)}\right)^2}\frac{\epsilon-\frac{4\pi^2\left(\sigma^{(1)}_i\right)^2}{c^2}}{\epsilon+\frac{4\pi^2\left(\sigma^{(1)}_i\right)^2}{c^2}}+\frac{\sigma_i^{(3)}}{\sigma_i^{(1)}}+\frac{\sigma_r^{(3)}}{\sigma_r^{(1)}}\right)\left|E_0\right|^2\right]\\
&=k_x^{i,{\rm lin}}\left[1+\left(\frac{c^2\chi^{(3)}}{8\pi\left(\sigma_i^{(1)}\right)^2}\frac{\epsilon-\frac{4\pi^2\left(\sigma^{(1)}_i\right)^2}{c^2}}{\epsilon+\frac{4\pi^2\left(\sigma^{(1)}_i\right)^2}{c^2}}+\frac{\sigma_i^{(3)}}{\sigma_i^{(1)}}+\frac{\sigma_r^{(3)}}{\sigma_r^{(1)}}\right)\left|E_0\right|^2\right].
\end{split}
\end{equation}
\end{subequations}
In the above, $k_x^{r,{\rm lin}}$ and $k_x^{i,{\rm lin}}$ denote the real and imaginary parts, respectively, of the SP wavenumber in the linear case; cf. Eq.~\eqref{eq:TE-disp-ex}.

By inspection of Eqs.~\eqref{eq:kx-TE-same-med}, we should add the following remarks.
Equation~\eqref{apr1} shows that the presence of the dielectric nonlinearity alone leads to an increase in the real part of the SP wavenumber (thus, a decrease of the SP wavelength), as Kerr media are predominantly focusing, $\chi^{(3)}>0$.
On the other hand, the effect of the nonlinearity of the 2D material is more complicated, as indicated by Eqs.~\eqref{eq:kx-TE-same-med}.
Specifically, the terms $\sigma_i^{(3)}/\sigma_i^{(1)}$ and $\sigma_r^{(3)}/\sigma_r^{(1)}$ can be positive or negative depending on the type of the conducting material and range of frequency, $\omega$.
In fact, if one takes into account the nonlinear behavior of the surface conductivity, it can be predicted that the wavelength and propagation length of a TE-polarized SP in the nonlinear regime can be larger or smaller than, or even nearly equal to, its linear counterpart.
The outcome of this comparison of course depends on the combined effect of the parameter values for the nonlinearities of the ambient dielectric and 2D material.

\section{TM-polarized surface plasmon}
\label{sec:TM}

In this section, we obtain the dispersion relation of the TM-polarized SP in the weakly dissipative regime. In this setting,
the electromagnetic field is written as
\begin{align*}
\bold{H}_j\left(x,z\right)&=\left(0, H_{jy}\left(z\right), 0\right)e^{ik_x x},\\
\bold{E}_j\left(x,z\right)&=\left(E_{jx}\left(z\right), 0, E_{jz}\left(z\right)\right)e^{ik_xx},\\
\bold{D}_j\left(x,z\right)&=\left(D_{jx}\left(z\right), 0, D_{jz}\left(z\right)\right)e^{ik_xx}\qquad (j=1,\,2),
\end{align*}
where $z>0$ for $j=1$ and $z<0$ for $j=2$. Hence, $\bold{E}_\parallel=\bold{e}_x E_{1x}=\bold{e}_x E_{2x}$ at $z=0$.
The substitution of the above expressions for $\bold{H}_j$, $\bold{E}_j$ and $\bold{D}_j$ into Eqs.~\eqref{m1} and~\eqref{m2} yields the following system of equations for the field components:
\begin{align*}
\frac{dH_{jy}}{dz}=\frac{i\omega}{c}D_{jx},\\
ik_xH_{jy}=-\frac{i\omega}{c}D_{jz},\\
\frac{dE_{jx}}{dz}-ik_xE_{jz}=\frac{i\omega}{c}H_{jy}.
\end{align*}
By eliminating the magnetic field, $H_{jy}$, from this system, we find that $E_{jx}$ and $E_{jz}$ obey the {\em coupled} equations
\begin{subequations}\label{eq:Exz-TM}
\begin{gather}
\label{eq:Exz-TM-1}
\frac{d^2E_{jx}}{dz^2}-ik_x\frac{dE_{jz}}{dz}=-k^2_j\left[1+\varepsilon_j\left(\left|E_{jx}\right|^2+\left|E_{jz}\right|^2\right)\right]E_{jx},\\
\label{eq:Exz-TM-2}
\frac{dE_{jx}}{dz}-ik_xE_{jz}=-\frac{k^2_j}{k_x}\left[1+\varepsilon_j\left(\left|E_{jx}\right|^2+\left|E_{jz}\right|^2\right)\right]E_{jz}.
\end{gather}
\end{subequations}
Recall that $k_j=\omega\sqrt{\epsilon_j}/c$ and $\varepsilon_j=4\pi\chi^{(3)}_j/\epsilon_j$ ($j=1,\,2$).

By using boundary conditions~\eqref{bc1} and~\eqref{bc2} along with constitutive law~\eqref{eq:js-nonl}, we obtain the relation
\begin{equation}
\label{bcc2}
\begin{split}
&\left\{k_1^2\left[1+\varepsilon_1\left(\left|E_{1x}\right|^2+\left|E_{1z}\right|^2\right)\right]E_{1z}\right.\\
&\left.-k_2^2\left[1+\varepsilon_2\left(\left|E_{2x}\right|^2+\left|E_{2z}\right|^2\right)\right]E_{2z}\right\}\biggl|_{z=0}=\frac{4\pi\omega k_x}{c^2}\sigma E_0,
\end{split}
\end{equation}
where $E_0=E_{1x}(0)=E_{2x}\left(0\right)$ is the value of the electric field on the 2D material sheet and
$\sigma=\sigma^{(1)}+\sigma^{(3)}\left|E_{0}\right|^2$.

In order to find the SP dispersion relation in view of Eq.~\eqref{bcc2}, we have to determine an additional relation between the electric field components, $E_{jx}$ and $E_{jz}$.
This relation can be extracted from Eqs.~\eqref{eq:Exz-TM} analytically in terms of the dielectric nonlinearities, $\varepsilon_j$.
For this purpose, we rewrite Eqs.~\eqref{eq:Exz-TM} in term of the variables $E_{jz}/E_{jx}$ and $\varepsilon_j|E_{jx}|^2$.
By analogy to the procedure in Sec.~\ref{sec:TE}, we treat each of these variables perturbatively:
We approximately write each one as a sum of the (zeroth-order) solution of the dissipation-free nonlinear system and a relatively small correction that accounts for dissipation and is linear in $k_x^i/k_x^r$.
In addition, we apply the condition $|k_x|\gg k_j$ $(j=1,2)$.
For details of this procedure, see Appendix~\ref{app:BX}.
As a result, we obtain
\begin{subequations}\label{eq:TM-pert-Ez}
\begin{gather}
\label{eq:TM-pert-E1z}
E_{1z}\approx E_{1x}\left[iF\left(\varepsilon_1\left|E_{1x}\right|^2\right)+\frac{k_x^i}{k_x^r}G\left(\varepsilon_1\left|E_{1x}\right|^2\right)\right],\\
\label{eq:TM-pert-E2z}
E_{2z}\approx-E_{2x}\left[iF\left(\varepsilon_2\left|E_{2x}\right|^2\right)+\frac{k_x^i}{k_x^r}G\left(\varepsilon_2\left|E_{2x}\right|^2\right)\right],
\end{gather}
\end{subequations}
where the real functions $F(a)$ and $G(a)$ are defined by
\begin{subequations}\label{eq:FG-def}
\begin{equation}\label{eq:F-def}
F\left(a\right)=\sqrt{\frac{\sqrt{4a^2+8a+1}-a-1}{3a}},
\end{equation}
\begin{align}\label{eq:G-def}
G\left(a\right)&=\frac{\left[1-3F(a)^2\right]^2}{F(a)^2-1}\left\{\frac{4+\pi}{16}-\frac{{\displaystyle \tanh^{-1}\frac{\sqrt{3}\left[F\left(a\right)-1\right]}{3F\left(a\right)-1}}}{4\sqrt{3}}\right.\notag \\
& \quad \left. -\frac{\tan^{-1} F\left(a\right)}{4}\right\}+\frac{F\left(a\right)\left[1-2F^2\left(a\right)\right]}{F\left(a\right)^2-1},\quad a>0.
\end{align}
\end{subequations}
Recall that we consider positive (focusing) Kerr nonlinearity of each dielectric;
thus, $a=\varepsilon_j|E_{jx}|^2$ is assumed to be positive.
We note in passing that the function $\tan^{-1}$ entering $G(a)$ is defined to have values in the interval $(-\pi/2,\pi/2)$.
As discussed below, by Eq.~\eqref{eq:F-def} $F(a)$ is properly bounded, consistent with the restrictions implied by the right-hand side of Eq.~\eqref{eq:G-def}.

It is of interest to comment on the significance of Eqs.~\eqref{eq:TM-pert-Ez}.
The functions $F(a)$ and $G(a)$ with $a=\varepsilon_j|E_{jx}|^2$ ($j=1,\,2$) express the cumulative (nonperturbative) effect of the Kerr nonlinearity on the requisite electric field components.
In particular, $F(a)$ is the non-dissipative contribution to $E_{jz}$ while $G(a)$ expresses the respective perturbation due to small enough dissipation.
To our knowledge, Eqs.~\eqref{eq:TM-pert-Ez} along with definitions~\eqref{eq:FG-def}, which combine the effect of small dissipation with the exact treatment of the Kerr nonlinearity, have not been reported previously.

For the sake of comparison, note that in the linear case (if $\varepsilon_j=0$) $F$ and $G$ must be replaced by the limiting values $F(0^+)=1$ and $G(0^+)=0$ (as $a$ approaches 0 from positive values); thus,  $E_{1z}\approx iE_{1x}$ and $E_{2z}\approx-iE_{2x}$ for weak dissipation.
Hence, the magnitudes of the electric field components are approximately equal to each other, $|E_{jz}|\approx |E_{jx}|$, while their phase difference is $\Delta\phi_j={\rm Arg}(E_{jz})-{\rm Arg}(E_{jx})\approx\left(-1\right)^{j+1}\pi/2$ ($j=1,\,2$). These properties imply that the SP in the linear regime is circularly polarized in the $xz$-plane.

By using Eqs.~\eqref{eq:TM-pert-Ez}, one can show that the dissipationless limit of the nonlinear problem, by which $k_x^i=0$, also corresponds to the phase difference $\Delta\phi_j\approx\left(-1\right)^{j+1}\pi/2$ between $E_{jz}$ and $E_{jx}$.
The magnitudes of these components are related to each other through the function $F$, viz., $\left|E_{jz}\right|\approx F(a)\left|E_{jx}\right|$ for $a=\varepsilon_j|E_{jx}|^2$, which describes the elliptization of the SP polarization due to the Kerr nonlinearity of the dielectric.
For this particular case, the dispersion relation of the SP is studied in \cite{hajian2014, hajian2016guided}.
Accordingly, for $k_x^i=0$, formulas~\eqref{eq:TM-pert-Ez} of our analysis reduce to the relation between $E_{1z}$ and $E_{1x}$ found in~\cite{hajian2014, hajian2016guided}.

Now let us further discuss the effect of dissipation ($k_x^i\neq0$).
By this effect, the SP polarization ellipse rotates.
This rotation is described by the phase difference $\Delta\phi_j$ (defined above) between the two electric field components, which depends on the magnitude of the electric field $|E_{jx}|$, viz.,
\begin{equation*}
\Delta\phi_j\approx\frac{\pi}{2}+\left(-1\right)^j\frac{k_x^i}{k_x^r}\left|\frac{G\left(\varepsilon_j\left|E_{jx}\right|^2\right)}{F\left(\varepsilon_j\left|E_{jx}\right|^2\right)}\right|.
\end{equation*}
Thus, the Kerr nonlinearity of the dielectric leads to both elliptization and rotation of SP polarization that are described by the parameters $F$ and $G$.

\begin{figure}[h]
\center{\includegraphics[width=1\linewidth]{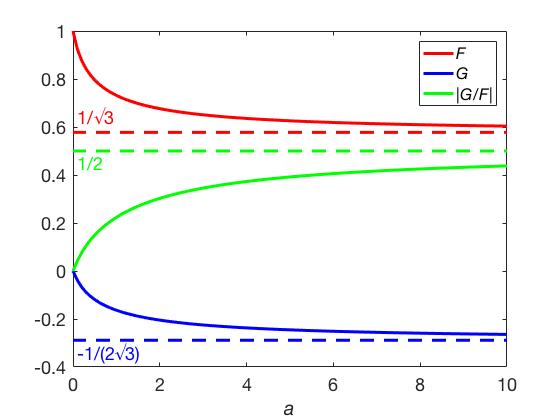}}
\centering{}\caption{(Color online) Plots of quantities $F(a)$, $G(a)$ and $|G(a)/F(a)|$ as a function of dielectric nonlinearity, $a=\varepsilon_j |E_{jx}|^2$.
These quantities characterize the elliptization of SP polarization through $F$ (red solid line); and the rotation of SP polarization through $G$ (blue solid line) and $|G/F|$ (green solid line).
The dashed lines correspond to horizontal asymptotes of $F$ (red line), $G$ (blue) and $|G/F|$ (green).}
\label{fig:ell}
\end{figure}

Figure~\ref{fig:ell} illustrates how the parameters $F$, $G$ and $\left|G/F\right|$, which control the SP polarization, depend on the nonlinearity of the dielectric (parameter $\varepsilon_j\left|E_{jx}\right|^2$) according to Eqs.~\eqref{eq:FG-def}.  Here, we assume that $\varepsilon_j$ is nonnegative.
We observe that  $F$, $G$ and $|G/F|$ are bounded, satisfying $1/\sqrt{3}<F\leq1$, $-1/(2\sqrt{3})<G\leq0$ and $\left|G/F\right|<1/2$.
Thus, we verify that the perturbation terms on the right-hand sides of
Eqs.~\eqref{eq:TM-pert-Ez}, which are proportional to $(k_x^i/k_x^r)G$ and express the dissipation effect, are indeed relatively small in the weakly dissipative regime ($k_x^i/k_x^r\ll 1$).

By inserting Eqs.~\eqref{eq:TM-pert-Ez} and~\eqref{eq:FG-def} into Eq.~\eqref{bcc2}, we obtain the desired dispersion relation in terms of the real and imaginary parts of the SP wavenumber, $k_x$.
The formulas are
\begin{subequations}\label{eq:TM-kx-ri}
\begin{widetext}
\begin{gather}
\label{drtm1}
\begin{split}
k_x^r&\approx \frac{\omega\sigma_i}{8\pi\left|\sigma\right|^2}\left[\tilde{\epsilon}_1\left\{F\left(\varepsilon_1\left|E_0\right|^2\right)-G\left(\varepsilon_1\left|E_0\right|^2\right)\right\}+\tilde{\epsilon}_2\left\{F\left(\varepsilon_2\left|E_0\right|^2\right)-G\left(\varepsilon_2\left|E_0\right|^2\right)\right\}\right]\\
&\times\left[1+\sqrt{1+4\frac{\left|\sigma\right|^2}{\sigma_i^2}\frac{\left\{\tilde{\epsilon}_1F\left(\varepsilon_1\left|E_0\right|^2\right)+\tilde{\epsilon}_2F\left(\varepsilon_2\left|E_0\right|^2\right)\right\}\left\{\tilde{\epsilon}_1G\left(\varepsilon_1\left|E_0\right|^2\right)+\tilde{\epsilon}_2G\left(\varepsilon_2\left|E_0\right|^2\right)\right\}}{\left[\tilde{\epsilon}_1\left\{F\left(\varepsilon_1\left|E_0\right|^2\right)-G\left(\varepsilon_1\left|E_0\right|^2\right)\right\}+\tilde{\epsilon}_2\left\{F\left(\varepsilon_2\left|E_0\right|^2\right)-G\left(\varepsilon_2\left|E_0\right|^2\right)\right\}\right]^2}}\right],
\end{split}\\
\label{drtm2}
\begin{split}
k_x^i\approx \frac{\omega\sigma_r}{4\pi\left|\sigma\right|^2}&\left[\tilde{\epsilon}_1F\left(\varepsilon_1\left|E_0\right|^2\right)+\tilde{\epsilon}_2F\left(\varepsilon_2\left|E_0\right|^2\right)\right]
\left[\frac{\left|\sigma\right|^2}{\sigma_r^2}-\frac{\sigma_i^2}{\sigma_r^2}\frac{\tilde{\epsilon}_1\left\{F\left(\varepsilon_1\left|E_0\right|^2\right)-G\left(\varepsilon_1\left|E_0\right|^2\right)\right\}+\tilde{\epsilon}_2\left\{F\left(\varepsilon_2\left|E_0\right|^2\right)-G\left(\varepsilon_2\left|E_0\right|^2\right)\right\}}{2\left\{\tilde{\epsilon}_1F\left(\varepsilon_1\left|E_0\right|^2\right)+\tilde{\epsilon}_2F\left(\varepsilon_2\left|E_0\right|^2\right)\right\}}\right.\\
&\left.\times\left(1+\sqrt{1+4\frac{\left|\sigma\right|^2}{\sigma_i^2}\frac{\left\{\tilde{\epsilon}_1F\left(\varepsilon_1\left|E_0\right|^2\right)+\tilde{\epsilon}_2F\left(\varepsilon_2\left|E_0\right|^2\right)\right\}\left\{\tilde{\epsilon}_1G\left(\varepsilon_1\left|E_0\right|^2\right)+\tilde{\epsilon}_2G\left(\varepsilon_2\left|E_0\right|^2\right)\right\}}{\left[\tilde{\epsilon}_1\left\{F\left(\varepsilon_1\left|E_0\right|^2\right)-G\left(\varepsilon_1\left|E_0\right|^2\right)\right\}+\tilde{\epsilon}_2\left\{F\left(\varepsilon_2\left|E_0\right|^2\right)-G\left(\varepsilon_2\left|E_0\right|^2\right)\right\}\right]^2}}\right)\right].
\end{split}
\end{gather}
\end{widetext}
\end{subequations}
In the above, we define $\tilde{\epsilon}_j=\epsilon_j[1+\varepsilon_j+\varepsilon_jF^2(\varepsilon_j|E_0|^2)]$ where $\varepsilon_j=4\pi\chi^{(3)}_j/\epsilon_j$ and $\sigma=\sigma^{(1)}+\sigma^{(3)}\left|E_0\right|^2$.
For the derivation of Eqs.~\eqref{eq:TM-kx-ri}, we assumed that $\sigma_i>0$, where $\sigma_i$ ($\sigma_r$) is the imaginary (real) part of $\sigma$.
Recall that in the linear regime the condition for the appearance of the TM-polarized SP is $\sigma_i^{(1)}>0$ (Sec.~\ref{subsec:plasmons}).
Equations~\eqref{eq:TM-kx-ri} hold in the quasi-electrostatic limit.

By inspection of Eqs.~\eqref{eq:TM-kx-ri}, it is evident that the SP wavenumber is nonlinear in $\chi^{(3)}_j$.
This dispersion relation cannot be obtained from the respective relation of the linear problem, Eq.~\eqref{eq:TM-quasi}, by replacement of $\epsilon_j$ with $\epsilon_j+4\pi\chi^{(3)}_j\left|E_0\right|^2$.
The reason for this complication is the nonlinear relation between the electric field components according
to Eqs.~\eqref{eq:TM-pert-Ez}:
This relation cannot  be approximated by $E_{jz}\approx\left(-1\right)^{j+1} iE_{jx}$, which characterizes the linear regime.
We note in passing that Eq.~\eqref{eq:TM-quasi} is recovered from Eq.~\eqref{eq:TM-kx-ri} by setting $\chi^{(3)}_j=0$ and $\sigma^{(3)}=0$.

If both media have the same dielectric permittivity, $\epsilon=\epsilon_1=\epsilon_2$, and third-order susceptibility, $\chi^{(3)}=\chi^{(3)}_1=\chi^{(3)}_2$, by Eqs.~\eqref{eq:TM-kx-ri} we find
\begin{subequations}\label{eq:kx-ri-same}
\begin{gather}
\label{k_x_r}
k_x^r\approx \frac{\omega\tilde{\epsilon}\left(F+\left|G\right|\right)}{4\pi}\frac{\sigma_i}{\left|\sigma\right|^2}
\left(1+\sqrt{1-\frac{\left|\sigma\right|^2}{\sigma_i^2}\frac{4\left|G/F\right|}{\left(1+\left|G/F\right|\right)^2}}\right),\\
\label{k_x_i}
k_x^i=\frac{\omega\tilde{\epsilon}F}{2\pi\sigma_r}\left[1-\frac{F+\left|G\right|}{2F}
\frac{\sigma_i^2}{\left|\sigma\right|^2}\left(1+\sqrt{1-\frac{\left|\sigma\right|^2}{\sigma_i^2}\frac{4\left|G/F\right|}{\left(1+\left|G/F\right|\right)^2}}\right)\right],
\end{gather}
\end{subequations}
where $F$ and $G$ are given by Eqs.~\eqref{eq:FG-def} with $a=\varepsilon |E_0|^2$, and $\varepsilon=4\pi \chi^{(3)}/\epsilon$.
 In Eqs.~\eqref{eq:kx-ri-same}, the expression under the square root is nonnegative, which entails the inequality
\begin{equation*}
	1-\frac{\left|\sigma\right|^2}{\sigma_i^2}\frac{4\left|G/F\right|}{\left(1+\left|G/F\right|\right)^2}\geq0.
\end{equation*}
 As we discussed above, $|G/F|<1/2$.
 Thus, Eqs.~\eqref{eq:kx-ri-same} hold for any positive Kerr nonlinearity, $\varepsilon\left|E_{0}\right|^2>0$, provided $\left|\sigma_r/\sigma_i\right|\leq1/\sqrt{8}$.
The last condition on $\sigma_r$ and $\sigma_i$ is satisfied within our approach since we restrict our analysis to the weakly dissipative regime, in which $|\sigma_r/\sigma_i|\ll 1$.

In fact, our assumption of weak dissipation, $|\sigma_r/\sigma_i|\ll 1$, allows us to simplify Eqs.~\eqref{eq:kx-ri-same} even further. By enforcing this regime explicitly, we obtain
\begin{subequations}\label{eq:kx-ri-same-wd}
\begin{gather}
\label{dr1}
k_x^r\approx\frac{\omega\epsilon\left[1+\varepsilon\left|E_0\right|^2+\varepsilon\left|E_{0}\right|^2F^2\right]F}{2\pi\sigma_i},\\
\label{dr2}
k_x^i\approx\frac{\omega\epsilon\left[1+\varepsilon\left|E_0\right|^2+\varepsilon\left|E_{0}\right|^2F^2\right]F}{2\pi}\frac{\sigma_r}{\sigma_i^2},
\end{gather}
\end{subequations}
where $F=F(a)$ is evaluated at $a=\varepsilon |E_0|^2$.  According to Eqs.~\eqref{eq:kx-ri-same-wd}, the ratio $k_x^i/k_x^r$ approximately becomes
\begin{equation}
\label{eq:damping-TM}
\frac{k_x^i}{k_x^r}\approx\frac{\sigma_r}{\sigma_i},
\end{equation}
and we have $|k_x^i/k_x^r|\approx|\sigma_r/\sigma_i|\ll 1$ which serves as a self-consistency check of our approximations for any $\varepsilon|E_0|^2>0$.
Notably, the damping of the TM-polarized SP, expressed by Eq.~\eqref{eq:damping-TM}, is independent of the nonlinearity of the dielectric to this leading order of our weak-dissipation approximation.

 By comparison of Eq.~\eqref{eq:dmp} with Eq.~\eqref{eq:damping-TM}, we observe that the relations for the damping of TE-polarized and TM-polarized SPs are similar. Recall that Eqs.~\eqref{eq:dmp} and \eqref{eq:damping-TM} correspond to different frequency regimes due to the mutually incompatible restrictions on $\sigma$. Hence, the ratio $k_x^i/k_x^r$ can be essentially different for the two polarizations since these frequency regimes can correspond to different transport mechanisms in the 2D material.  Later on, we discuss the effect on SP of the nonlinearity of the surface conductivity of the 2D material for the particular case of graphene (see Sec.~\ref{sec:discussion}).

By using Eqs.~\eqref{eq:kx-ri-same-wd} and taking into account the property $|\sigma|^2\approx\sigma_i^2$ and the definition $\sigma=\sigma^{(1)}+\sigma^{(3)}\left|E_0\right|^2$, we derive the following expression for the (complex) SP wavenumber:
\begin{equation}
\label{eq:kx-cmpx}
k_x\approx\frac{\omega\epsilon\left[1+\varepsilon\left|E_0\right|^2+\varepsilon\left|E_{0}\right|^2F^2\right]F}{2\pi\left(-i\right)\left(\sigma^{(1)}+\sigma^{(3)}\left|E_0\right|^2\right)}.
\end{equation}
To our knowledge, dispersion relation~\eqref{eq:kx-cmpx} has not been previously reported.
It describes the nonperturbative effect of dielectric and graphene nonlinearities on the SP wavenumber.
Evidently, this dispersion relation does not depend on $G$.
This quantity, $G$, only impacts the higher-order correction terms which are of the order of $(\sigma_r/\sigma_i)^2$ in our weak-dissipation approximation scheme.

According to Eq.~\eqref{eq:kx-cmpx}, the nonlinearity of the dielectric, expressed by the {\em positive} third-order susceptibility $\chi^{(3)}$, causes an {\em increase} to both the real and imaginary parts of the SP wavenumber in the present case of TM polarization.
In contrast, in regard to the nonlinearity of the surface conductivity, the sign of $\sigma^{(3)}$ depends on the particular 2D material and operating frequency, $\omega$, of the incident electromagnetic field.
A more detailed discussion on this issue for graphene is provided in Sec.~\ref{sec:discussion}.

For weak nonlinearities of the dielectric medium, if $\varepsilon=4\pi\chi^{(3)}/\epsilon\ll 1$, we can show that $F(\varepsilon|E_0|^2)\approx 1-\varepsilon|E_{0}|^2$ with an error of the order of $\varepsilon^2|E_{0}|^4$, while $G(\varepsilon|E_0|^2)$ is of the order of $\varepsilon^2\left|E_{0}\right|^4$.
Assuming that the 2D material nonlinearity is also small, i.e., $|\sigma^{(3)}/\sigma^{(1)}|\ll 1$, we can write dispersion relation~\eqref{eq:kx-cmpx} as
\begin{equation*}
\begin{split}
k_x&\approx\frac{\omega\epsilon}{2\pi\left(-i\sigma^{(1)}\right)}\left(1+\left[\frac{4\pi\chi^{(3)}}{\epsilon}-\frac{\sigma^{(3)}}{\sigma^{(1)}}\right]\left|E_{0}\right|^2\right)\\
&=k_x^{\rm lin}\left(1+\left[\frac{4\pi\chi^{(3)}}{\epsilon}-\frac{\sigma^{(3)}}{\sigma^{(1)}}\right]\left|E_{0}\right|^2\right).
\end{split}
\end{equation*}
Here, $k_x^{\rm lin}$ denotes the wavenumber of the TM-polarized SP in the linear regime, and is given by Eq.~\eqref{eq:TM-quasi}.
The above dispersion relation is in agreement with the corresponding one obtained using perturbations of Maxwell's equations for small nonlinearities in~\cite{gorbach2013}.

\section{Discussion}
\label{sec:discussion}

The analytical results obtained thus far aim to describe generally the dispersion of SPs in a wide family of nonlinear isotropic materials  characterized by inversion symmetry.
In this section, we discuss in more detail the effect of material nonlinearities on the wavelength and propagation distance of TM-polarized SPs.
 We also compare these features to those of TE-polarized SPs.
For definiteness, in our discussion we place some emphasis on the case when the 2D material is doped graphene.
This is a well-studied 2D material.
For example, the third-order conductivity of this material has been the subject of extensive investigations~\cite{mikhailov2007, wright2009, cheng2014third, cheng2015third, mikhailov2016}.

First, let us consider the simplified setting with a linear 2D material lying in a nonlinear dielectric medium, thus setting $\sigma^{(3)}=0$ (along with $\varepsilon_1=\varepsilon_2=\varepsilon$).
By using dispersion relation~\eqref{eq:kx-cmpx} for TM-polarized SPs , we obtain the formula
\begin{equation}
\label{eq:shift}
\frac{k_x}{k_x^{\rm lin}}\approx \left[1+\varepsilon\left|E_0\right|^2+\varepsilon\left|E_{0}\right|^2F^2\left(\varepsilon\left|E_0\right|^2\right)\right]F\left(\varepsilon\left|E_0\right|^2\right),
\end{equation}
where $k_x^{\rm lin}$ and $F(a)$ are given by Eqs.~\eqref{eq:TM-quasi} and~\eqref{eq:F-def}, respectively.
Note that the ratio $k_x/k_x^{\rm lin}$ does not depend on frequency for a given value of the nonlinear parameter $\varepsilon |E_0|^2$.
We repeat at the risk of redundancy that the Kerr nonlinearity of the dielectric is assumed to be focusing, so that $\varepsilon>0$.

\begin{figure}[h]
\center{\includegraphics[width=1\linewidth]{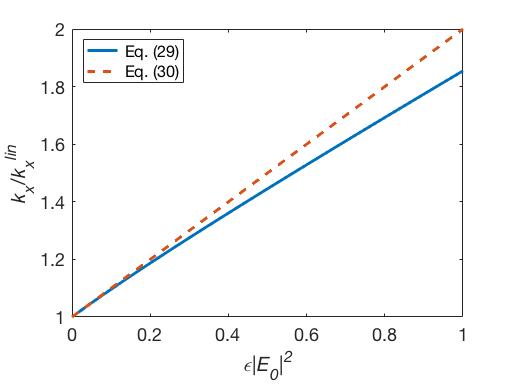}}
\centering{}\caption{(Color online) Wavenumber, $k_x$, of TM-polarized SP scaled by its linear counterpart, $k_x^{\rm lin}$, versus nonlinearity, $\varepsilon\left|E_0\right|^2$, of the ambient dielectric, with linear 2D material ($\sigma^{(3)}=0$). Two distinct dispersion relations are used: Equation~\eqref{eq:shift} from systematic treatment of Maxwell's equations (blue solid line); and Eq.~\eqref{eq:TM-kx-naive} from naive replacement of nonlinear dielectric permittivity in dispersion relation of linear regime (orange dashed line).}.
\label{fig:TM-vs-naive}
\end{figure}

Now consider instead the dispersion relation
\begin{equation}\label{eq:TM-kx-naive}
	\frac{k_x}{k_x^{\rm lin}}\approx 1+\varepsilon\left|E_0\right|^2,
\end{equation}
which results from naively replacing the dielectric permittivity by its nonlinear version in the dispersion relation of the linear regime, Eq.~\eqref{eq:TM-quasi}.
Figure~\ref{fig:TM-vs-naive} displays the comparison between dispersion relation~\eqref{eq:shift} and its naive yet simpler counterpart~\eqref{eq:TM-kx-naive}.
It is evident that Eq.~\eqref{eq:TM-kx-naive} approximates the wavenumber, $k_x$, of a
TM-polarized SP reasonably well for $\varepsilon |E_0|^2\lesssim 0.2$.
However, it is evident that the naive prediction overestimates $k_x$ for large enough values of $\varepsilon\left|E_0\right|^2$.

Next, we include in the discussion  of TM-polarized SPs  the nonlinearity of the conductivity of the 2D material.
To better understand the ensuing joint effect of the nonlinearities on the SP dispersion relation, we scale Eqs.~\eqref{dr1} and~\eqref{dr2} by the real part, $k_x^{{\rm lin},r}$, and imaginary part, $k_x^{{\rm lin},i}$, of the SP wavenumber, $k_x^{\rm lin}$, in the linear regime, respectively.
The resulting equations read
\begin{subequations}\label{eq:TM-realim-3cond}
\begin{gather}
\frac{k_x^r}{k_x^{{\rm lin},r}}\approx \frac{\left(1+\varepsilon\left|E_0\right|^2+\varepsilon\left|E_{0}\right|^2F^2\right)F}{1+\frac{\sigma^{(3)}_i}{\sigma^{(1)}_i}\left|E_{0}\right|^2},\label{eq:TM-realim-3cond-1}\\
\frac{k_x^i}{k_x^{{\rm lin},i}}\approx \frac{\left(1+\varepsilon\left|E_0\right|^2+\varepsilon\left|E_{0}\right|^2F^2\right)F}{\left(1+\frac{\sigma^{(3)}_i}{\sigma^{(1)}_i}\left|E_{0}\right|^2\right)^2}\left(1+\frac{\sigma^{(3)}_r}{\sigma^{(1)}_r}\left|E_{0}\right|^2\right). \label{eq:TM-realim-3cond-2}
\end{gather}
\end{subequations}

Note that the ratio $\sigma^{(3)}_i|E_{0}|^2/\sigma^{(1)}_i$ is present in both of the above formulas.
In contrast, $\sigma^{(3)}_r|E_{0}|^2/\sigma^{(1)}_r$, which pertains to dissipation in the 2D material, enters only the formula for $k_x^i/k_x^{{\rm lin},i}$.
The quantity $\sigma^{(3)}_i|E_{0}|^2/\sigma^{(1)}_i$ can, in principle, be negative or positive (depending on the specific material and frequency).
In doped graphene, $\sigma^{(3)}_i(\omega)$ has a negative sign in a suitable (THz) frequency range~\cite{cheng2014third, cheng2015third, mikhailov2016}.

\begin{figure}[h]
\center{\includegraphics[width=1\linewidth]{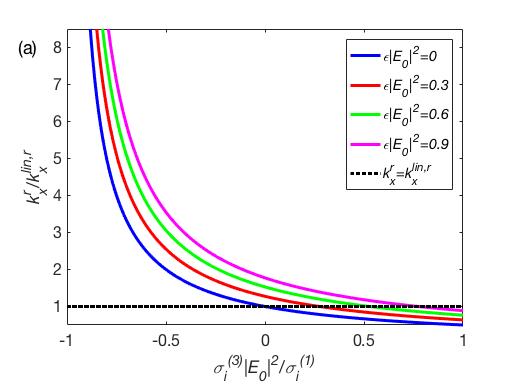}}
\center{\includegraphics[width=1\linewidth]{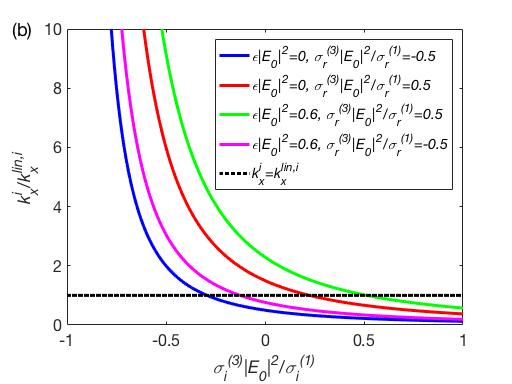}}
\centering{}\caption{(Color online) Scaled wavenumber of TM-polarized SP as a function of scaled third-order conductivity, $\sigma^{(3)}_i|E_{0}|^2/\sigma^{(1)}_i$, of 2D material for different values of nonlinearity, $\epsilon|E_0|^2$, of ambient dielectric medium.
Top panel [(a)]: $k_x^r/k_x^{{\rm lin},r}$ versus $\sigma^{(3)}_i|E_{0}|^2/\sigma^{(1)}_i$ by Eq.~\eqref{eq:TM-realim-3cond-1}.
The plots are independent of $\sigma_r^{(3)}|E_0|^2/\sigma^{(1)}_r$.
Bottom panel [(b)]: $k_x^i/k_x^{{\rm lin},i}$ versus $\sigma^{(3)}_i|E_{0}|^2/\sigma^{(1)}_i$ by Eq.~\eqref{eq:TM-realim-3cond-2}.
The quantity $k_x^{\rm lin}$ is the corresponding SP wavenumber in linear regime.
The dotted black line corresponds to the relation $k_x/k_x^{\rm lin}=1$ for real parts $k_x^r$ and $k_x^{{\rm lin},r}$ (a) and imaginary parts $k_x^i$ and $k_x^{{\rm lin},i}$(b).}
\label{fig:TM-vs-3cond}
\end{figure}

Figure~\ref{fig:TM-vs-3cond} illustrates the dependence of quantities $k_x^r/k_x^{{\rm lin},r}$ and $k_x^i/k_x^{{\rm lin},i}$ on the scaled nonlinearity $\sigma^{(3)}_i|E_{0}|^2/\sigma^{(1)}_i$, of the 2D material according to Eqs.~\eqref{eq:TM-realim-3cond}, for different values of the (positive) Kerr nonlinearity, $\varepsilon |E_0|^2$.  Note that the nonlinear parameters $\sigma_i^{(3)}|E_0|^2/\sigma_i^{(1)}$ and $\sigma_r^{(3)}|E_0|^2/\sigma_r^{(1)}$ are material-specific and, in principle, frequency dependent. However, by Eqs.~\eqref{eq:TM-realim-3cond}, the ratio between the SP wavenumber, $k_x$, and its linear counterpart, $k_x^{\rm lin}$, does not depend on frequency explicitly. Thus, without specifying the material, we can consider the ratios $k_x^r/k_x^{{\rm lin},r}$ and $k_x^i/k_x^{{\rm lin},i}$ as functions of $\sigma_i^{(3)}|E_0|^2/\sigma_i^{(1)}$ and $\sigma_r^{(3)}|E_0|^2/\sigma_r^{(1)}$. For doped graphene, we assume that the parameters $\sigma_r^{(1)}$, $\sigma_i^{(1)}$, $\sigma_r^{(3)}$ and $\sigma_i^{(3)}$ are dependent on frequency according to Eqs.~\eqref{eq:Drude}, \eqref{eq:Drude3}.

As seen in Fig.~\ref{fig:TM-vs-3cond}a, which depicts Eq.~\eqref{eq:TM-realim-3cond-1}; if $\sigma^{(3)}_i|E_{0}|^2/\sigma^{(1)}_i<0$, then the real part, $k_x^r$, of the SP wavenumber is larger than the corresponding quantity, $k_x^{{\rm lin},r}$, of the linear case regardless of the magnitude of $\varepsilon |E_0|^2$.
In fact, we notice that a negative third-order conductivity of the 2D material further improves the fine scale of the TM-polarized SP with increasing $\varepsilon |E_0|^2$.
In contrast, $k_x^r$ exhibits a more complicated behavior if $\sigma^{(3)}_i|E_{0}|^2/\sigma^{(1)}_i>0$. In this regime of positive nonlinearity of the 2D material, $k_x^r$ can be smaller than its counterpart of the linear regime if $\varepsilon|E_0|^2$ is sufficiently weak.
On the other hand, by Eq.~\eqref{eq:TM-realim-3cond-2}, the effect of the conductivity nonlinearity of the 2D material on $k_x^i/k_x^{{\rm lin},i}$ is determined by the value of $\sigma^{(3)}_i|E_{0}|^2/\sigma^{(1)}_i$ relative to $\sigma^{(3)}_r|E_{0}|^2/\sigma^{(1)}_r$.
This effect is depicted in Fig.~\ref{fig:TM-vs-3cond}b.

Interestingly, in Fig.~\ref{fig:TM-vs-3cond} we notice that there are values for $\sigma^{(3)}_i|E_{0}|^2/\sigma^{(1)}_i$, $\sigma^{(3)}_r|E_{0}|^2/\sigma^{(1)}_r$ and $\varepsilon|E_0|^2$ such that the resulting $k_x$ becomes nearly equal to the corresponding quantity, $k_x^{{\rm lin}}$, of the linear regime.
The parameter values are: $\sigma^{(3)}_r|E_{0}|^2/\sigma^{(1)}_r\approx0.5$, $\varepsilon|E_0|^2\approx0.6$ and $\sigma^{(3)}_i|E_{0}|^2/\sigma^{(1)}_i\approx0.5$.
Thus, the respective dielectric and 2D material nonlinearities may possibly balance each other out to cause SP dispersion similar to that through linear media.

\begin{figure}[h]
\center{\includegraphics[width=1\linewidth]{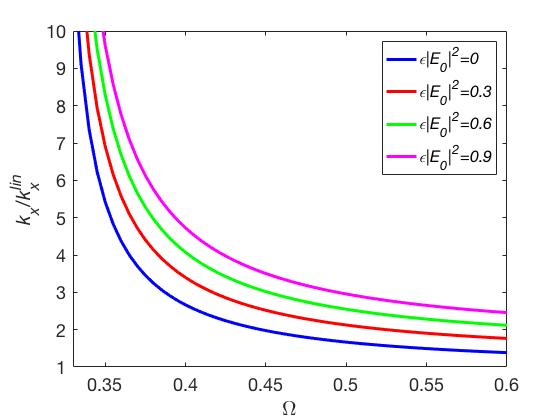}}
\centering{}\caption{(Color online) Wavenumber, $k_x$, of TM-polarized SP in graphene relative to wavenumber, $k_x^{\rm lin}$, of linear regime as a function of scaled frequency, $\Omega$, according to Eq.~\eqref{eq:TM-Drude}. The plots correspond to different values of nonlinearity parameter, $\varepsilon|E_0|^2$, of ambient dielectric. In all graphs, we use the values $\Gamma=0.01$ and $\sigma^{(3)}_0|E_{0}|^2/\sigma^{(1)}_0=0.1$.}
\label{fig:4}
\end{figure}


Next, consider the case of graphene with surface conductivity described by Eqs.~\eqref{eq:Drude} and~\eqref{eq:Drude3}.
Accordingly, dispersion relation~\eqref{eq:kx-cmpx}  for TM-polarized SPs  is expressed as
\begin{equation}
\label{eq:TM-Drude}
\frac{k_x}{k_x^{\rm lin}}=\frac{\left(1+\varepsilon\left|E_0\right|^2+\varepsilon\left|E_{0}\right|^2F^2\right)F}{1-\left(\sigma^{(3)}_0\left|E_{0}\right|^2/\sigma^{(1)}_0\right)\left(\Omega^2+\Gamma^2\right)^{-1}},
\end{equation}
where $F=F(\varepsilon|E_0|^2)$.
The above formula explicitly shows the frequency ($\Omega$-) dependence of the SP wavenumber relative to the linear case.
Equation~\eqref{eq:TM-Drude} is valid when the denominator is positive, $1-\left(\sigma^{(3)}_0\left|E_{0}\right|^2/\sigma^{(1)}_0\right)\left(\Omega^2+\Gamma^2\right)^{-1}>0$.
This condition results from the perturbation model for the graphene conductivity \cite{mikhailov2016}, which implies that $\left|\sigma^{(1)}\right|>\left|\sigma^{(3)}\right|\left|E_0\right|^2$.

In Fig.~\ref{fig:4}, we plot $k_x/k_x^{\rm lin}$ as a function of the scaled frequency $\Omega$ for $\Gamma=0.01$ and $\sigma^{(3)}_0|E_{0}|^2|/\sigma^{(1)}_0=0.1$, which corresponds to the values $E_F\approx0.1$ eV and $E_0\approx45$ kV/cm.
We observe that the nonlinearities of the graphene conductivity and ambient dielectric both cause an increase of the real and imaginary parts of the SP wavenumber relative to the corresponding quantities of the linear regime.
Notably, the damping, $k_x^i/k_x^r$, of the TM-polarized SP  is the same in the linear and nonlinear regimes at fixed $\Omega$. Indeed, by Eq.~\eqref{eq:TM-Drude} we obtain
\begin{equation}
\label{eq:damping-graphene-TM}
	\frac{k_x^i}{k_x^r}=\frac{k_x^{{\rm lin},i}}{k_x^{{\rm lin},r}}=\frac{\Gamma}{\Omega}.
\end{equation}

\begin{figure}[h]
\center{\includegraphics[width=1\linewidth]{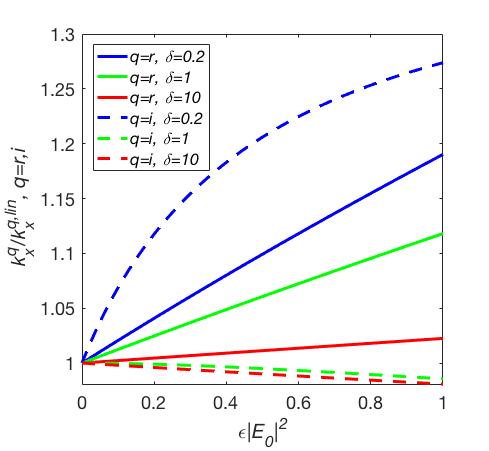}}
\centering{}\caption{(Color online)  Real (solid lines) and imaginary (dashed lines) parts of the wavenumber, $k_x^r$ and $k_x^i$, of the TE-polarized SP scaled by their linear counterparts, $k_x^{\rm r,lin}$ and $k_x^{\rm i,lin}$, versus nonlinearity, $\varepsilon\left|E_0\right|^2$, of the ambient dielectric. The 2D material is linear ($\sigma^{(3)}=0$) for different values of the parameter $\delta=4\pi^2(\sigma^{(1)}_i)^2/(c^2\epsilon)$: $\delta=0.2, 1, 10$.}.
\label{fig:TE-kxir}
\end{figure}

It is of interest to compare the TM dispersion relation~\eqref{eq:shift} with its TE counterpart obtained from Eqs.~\eqref{eq:kx-TE}; see Fig.~\ref{fig:TE-kxir}. Note that in the case of a TE-polarized SP, the dispersion relation is described by two parameters, $k_x^{\rm r}/k_x^{\rm r, lin}$ and $k_x^{\rm i}/k_x^{\rm i, lin}$. By tuning the value of the 2D material conductivity, $\sigma^{(1)}_i$, and, hence, the ratio $\delta=4\pi^2(\sigma^{(1)}_i)^2/(c^2\epsilon)$ one can, in principle, increase $k_x^{\rm r}/k_x^{\rm r, lin}$ and decrease $k_x^{\rm i}/k_x^{\rm i, lin}$ simultaneously.

\begin{figure}[h]
\center{\includegraphics[width=1\linewidth]{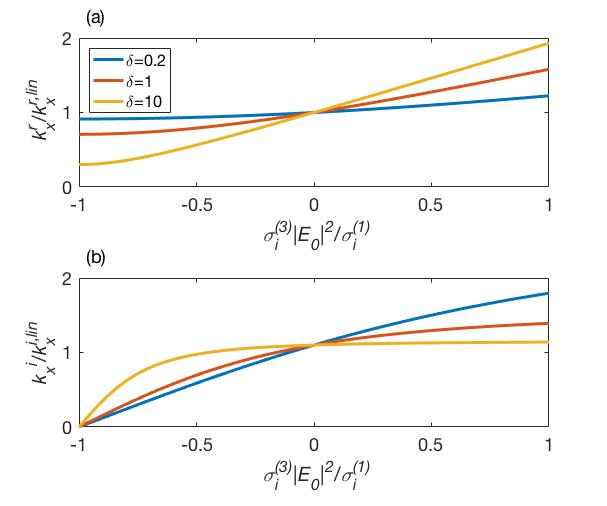}}
\centering{}\caption{(Color online)  Scaled wavenumber of the TE-polarized SP as a function of scaled third-order conductivity, $\sigma^{(3)}_i|E_{0}|^2/\sigma^{(1)}_i$, of a 2D material for different values of $\delta=4\pi^2(\sigma^{(1)}_i)^2/(c^2\epsilon)$.
Top panel [(a)]: $k_x^r/k_x^{{\rm lin},r}$ versus $\sigma^{(3)}_i|E_{0}|^2/\sigma^{(1)}_i$ by Eq.~\eqref{kxr}.
Bottom panel [(b)]: $k_x^i/k_x^{{\rm lin},i}$ versus $\sigma^{(3)}_i|E_{0}|^2/\sigma^{(1)}_i$ by Eq.~\eqref{kxi}.
For both (a) and (b), the values of $\epsilon|E_0|^2$ and $\sigma^{(3)}_r|E_{0}|^2/\sigma^{(1)}_r$ are $0$ and $0.1$ respectively.}
\label{fig:TEsigma}
\end{figure}

Recall that a negative third-order conductivity of the 2D material further improves the fine scale of the TM-polarized SP (Fig.~\ref{fig:TM-vs-3cond}).
In contrast, in the TE case, a positive third-order conductivity of the 2D material increases the real part of the SP wavenumber compared to the linear regime; see Fig.~\ref{fig:TEsigma}.

\begin{figure}[h]
\center{\includegraphics[width=1\linewidth]{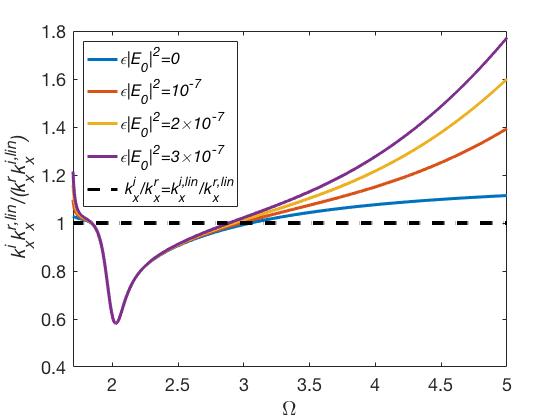}}
\centering{}\caption{(Color online)  Scaled damping of the TE-polarized SP in doped graphene as a function of scaled frequency, $\Omega$, according to Eqs.~\eqref{eq:kx-TE}.
The plots correspond to different values of the nonlinearity parameter, $\varepsilon|E_0|^2$, of the ambient dielectric. In all graphs, we use the values $\Gamma=0.1$ and $\sigma^{(3)}_0|E_{0}|^2/\sigma^{(1)}_0=0.01$.}
\label{fig:TEdamping}
\end{figure}

We have shown that the damping of TM-polarized SPs in doped graphene in the nonlinear weakly dissipative regime is the same as in the linear regime: see Eq.~\eqref{eq:damping-graphene-TM}. Notably, for the TE-polarized SP in graphene the corresponding damping is comparable to the damping in the linear regime for a narrow range of frequencies expressed by the nondimensional parameter $\Omega$; see Fig.~\ref{fig:TEdamping}. Interestingly, the nonlinearity parameter, $\varepsilon|E_0|^2$, of the ambient dielectric affects the damping of TE modes in a weakly dissipative regime mostly in the frequency region $\Omega>3$. In contrast, the damping of TM-polarized SPs does not depend on $\varepsilon|E_0|^2$ (Eq.~\eqref{eq:damping-graphene-TM}).
Note that the frequency range in Fig.~\ref{fig:TEdamping} is $1.7<\Omega<5$ which corresponds to the negative value of the imaginary part of graphene conductivity, $\sigma_i<0$.
In Fig.~\ref{fig:TEdamping}, we use the expressions for the linear and third-order conductivity of graphene, $\sigma^{(1)}\left(\Omega\right)$ and $\sigma^{(3)}\left(\Omega\right)$, derived in \cite{mikhailov2016}.

%
%
\section{Conclusion}
\label{sec:conclusion}

In this paper, we analytically derived the dispersion relations for TM- and TE-polarized SPs on nonlinear 2D materials  with inversion symmetry  that form boundaries between two semi-infinite, Kerr dielectric media.
In our approach, we relaxed some of the commonly used assumptions of previous works.
For instance, we took into account the small dissipation in the 2D material.
In addition, we determined the exact contributions of nonlinearities of the dielectric and 2D material to both the wavelength and propagation distance of the SP.

We find that the wavelength and propagation distance of SPs decrease when the nonlinearity of the dielectric is included.
In contrast, the effect of the nonlinearity of the 2D material on the dispersion relations depends on the signs of both the real and imaginary parts of the third-order conductivity, $\sigma^{(3)}$.
In the case of doped graphene, the $\sigma^{(3)}$ in the THz frequency range causes a decrease of the TM-polarized SP wavelength and propagation distance.

Our analysis admits several extensions, such as to nonlinear effects related to frequency conversion, the influence of the spatial and temporal shape of a source field, moderate dissipation in the 2D material and 2D materials with broken inversion symmetry.
It will be worthwhile for a future effort to study the properties of SPs in the nonlinear regime in 2D materials other than graphene, such as black phosphorus and MoS$_2,$ once their nonlinear conductivities as a function of frequency are calculated.


\acknowledgements

We acknowledge support by ARO MURI Award No. W911NF-14-0247 (V.A., M.L., D.M.) and NSF Grant No. DMS-1412769 (D.M.).
D.M. acknowledges the support of the Institute of Mathematics and its Applications for several visits, and we acknowledge
discussions with Prof. Tony Low and participants in the IMA Workshop on Theory and Computation for Transport Properties in 2D Materials.

\begin{appendix}

\section{On the electric field for TE polarization}
\label{app:AX}
In this appendix, we derive Eqs.~\eqref{eq:dEydz}. The starting point is to write the electric field component as $E_{jy}=|E_{jy}|e^{i\phi_j}$ ($j=1,\,2$). In the weakly dissipative regime, the magnitude, $|E_{jy}|$, and phase, $\phi_j$, of $E_{jy}$ can be expanded as
\begin{gather*}
|E_{jy}|\approx|E_{jy}|^{(0)}+\gamma |E_{jy}|^{(1)},\\
\phi_j\approx\phi_j^{(0)}+\gamma\phi_j^{(1)};\ \quad \gamma=k_x^i/k_x^r.
\end{gather*}
Note that $|E_{jy}|^{(0)}$, $|E_{jy}|^{(1)}$, $\phi_j^{(0)}$ and $\phi_j^{(1)}$ do not depend on $\gamma$.

By using Eqs.~\eqref{eq:Ey-mag-phase} and separating the real and imaginary parts in the corresponding expressions, we obtain the following equations for $|E_{jy}|^{(0)}$, $|E_{jy}|^{(1)}$, $\phi_j^{(0)}$ and $\phi_j^{(1)}$ ($j=1,\,2$):
\begin{widetext}
\begin{subequations}
\begin{gather}
\label{a1}
\frac{d^2|E_{jy}|^{(0)}}{dz^2}-|E_{jy}|^{(0)}\left(\frac{d\phi_j^{(0)}}{dz}\right)^2=\left(\left(k_x^r\right)^2-k_j^2-k_j^2\varepsilon_j\left(|E_{jy}|^{(0)}\right)^2\right)|E_{jy}|^{(0)},\\
\label{a2}
2\frac{d\left|E_{jy}\right|^{(0)}}{dz}\frac{d\phi_j^{(0)}}{dz}+\left|E_{jy}\right|^{(0)}\frac{d^2\phi_j^{(0)}}{dz^2}=0,\\
\label{a3}
\frac{d^2\left|E_{jy}\right|^{(1)}}{dz^2}-\left|E_{jy}\right|^{(1)}\left(\frac{d\phi_j^{(0)}}{dz}\right)^2-2\left|E_{jy}\right|^{(0)}\frac{d\phi_j^{(0)}}{dz}\frac{d\phi_j^{(1)}}{dz}=\left(\left(k_x^r\right)^2-k_j^2\right)\left|E_{jy}\right|^{(1)}-3k_j^2\varepsilon_j\left(\left|E_{jy}\right|^{(0)}\right)^2\left|E_{jy}\right|^{(1)},\\
\label{a4}
2\frac{d\left|E_{jy}\right|^{(1)}}{dz}\frac{d\phi_j^{(0)}}{dz}+2\frac{d\left|E_{jy}\right|^{(0)}}{dz}\frac{d\phi_j^{(1)}}{dz}+\left|E_{jy}\right|^{(0)}\frac{d^2\phi_j^{(1)}}{dz^2}+\left|E_{jy}\right|^{(1)}\frac{d^2\phi_j^{(0)}}{dz^2}=2\left(k_x^r\right)^2\left|E_{jy}\right|^{(0)}.
\end{gather}
\end{subequations}
\end{widetext}

Equations~\eqref{a1} and~\eqref{a2} describe the lossless system~\cite{bludov2014, wu2017, qasymeh2017}.
Their solution consists of $d\phi_j^{(0)}/dz=0$ along with the formula
\begin{align}
\label{a5}
\left|E_{jy}\right|^{(0)}&=\sqrt{\frac{2\left(\left(k_x^r\right)^2-k_j^2\right)}{k_j^2\varepsilon_j}}\notag\\
&\qquad \times \sech\left(\left(-1\right)^{j+1}\sqrt{\left(k_x^r\right)^2-k_j^2}z+C_1\right),
\end{align}
where $C_1$ is a constant.
The $z$-derivative of Eq.~\eqref{a5} is
\begin{equation*}
\frac{d\left|E_{jy}\right|^{(0)}}{dz}=\left(-1\right)^j\sqrt{\left(k_x^r\right)^2-k_j^2-\frac{1}{2}k_j^2\varepsilon_j\left(\left|E_{jy}\right|^{(0)}\right)^2}\left|E_{jy}\right|^{(0)}.
\end{equation*}
Accordingly, the suitable solution of Eq.~\eqref{a3} is $\left|E_{jy}\right|^{(1)}=0$.
Hence, Eq.~\eqref{a4} becomes
\begin{align}
\label{a8}
\frac{d^2\phi_j^{(1)}}{dz^2}&=2\left(k_x^r\right)^2+2\left(-1\right)^{j+1}\notag\\&\quad\times \sqrt{\left(k_x^r\right)^2-k_j^2-\frac{1}{2}k_j^2\varepsilon_j\left(\left|E_{jy}\right|^{(0)}\right)^2}\frac{d\phi_j^{(1)}}{dz}.
\end{align}

By use of Eqs.~\eqref{a5} and~\eqref{a8}, we obtain the formula
\begin{widetext}
\begin{align*}
\frac{d\phi_j^{(1)}}{dz}&=\frac{2\left(k_x^r\right)^2}{k_j^2\varepsilon_j\left(\left|E_{jy}\right|^{(0)}\right)^2}
\left\{C_2\cosh^2\left(\left(-1\right)^{j+1}\sqrt{\left(k_x^r\right)^2-k_j^2}z+C_1\right)+\sinh\left(\left(-1\right)^{j+1}\sqrt{\left(k_x^r\right)^2-k_j^2}z+C_1\right)\right.\notag\\
&\qquad \times \left.\cosh\left(\left(-1\right)^{j+1}\sqrt{\left(k_x^r\right)^2-k_j^2}z+C_1\right)\right\}.
\end{align*}
\end{widetext}
Hence, the derivative of $\phi_j^{(1)}$ at $z=0$ can be expressed in terms of $|E_0|$, the value of $|E_{jy}|^{(0)}$ at $z=0$, as follows:

\begin{align*}
\left.\frac{d\phi_j^{(1)}}{dz}\right|_{z=0}&=\left(-1\right)^{j+1}\frac{4\left(k_x^r\right)^2}{k_j^2\varepsilon_j\left|E_0\right|^2}\left(\sqrt{\left(k_x^r\right)^2-k_j^2-\frac{1}{2}k_j^2\varepsilon_j\left|E_0\right|^2}\right.\notag\\
& \left. -\sqrt{\left(k_x^r\right)^2-k_j^2}\right).
\end{align*}
Therefore, the $z$-derivative of the electric field is
\begin{align*}
\left.\frac{dE_{jy}}{dz}\right|_{z=0}&=\left(-1\right)^{j}\sqrt{\left(k_x^r\right)^2-k_j^2-\frac{1}{2}k_j^2\varepsilon_j\left|E_0\right|^2}E_0 \,\left\{1 \right. \notag\\
& \left. -\frac{4ik_x^rk_x^i}{k_j^2\varepsilon_j\left|E_0\right|^2}\left(1-\sqrt{\frac{\left(k_x^r\right)^2-k_j^2}{\left(k_x^r\right)^2-k_j^2-\frac{1}{2}k_j^2\varepsilon_j\left|E_0\right|^2}}\right)\right\}.
\end{align*}

\section{On the electric field for TM polarization}
\label{app:BX}
In this appendix, we derive Eqs.~\eqref{eq:TM-pert-Ez} and~\eqref{eq:FG-def}.
First, we find $d^2E_{jx}/dz^2$ by differentiating Eq.~\eqref{eq:Exz-TM-2}, and substitute the result into Eq.~\eqref{eq:Exz-TM-1}.
Consequently, we obtain the following system of differential equations:
\begin{widetext}
\begin{gather*}
\frac{dE_{jz}}{dz}=-ik_xE_{jx}+i\left(k_x+k_x^*\right)\frac{\varepsilon\left(\left|E_{jz}\right|^2E_{jx}-E_{jz}^2E_{jx}^*\right)}{1+\varepsilon_j\left(\left|E_{jx}\right|^2+3\left|E_{jz}\right|^2\right)}
+i\varepsilon_jk_j^2\frac{1+\varepsilon_j\left(\left|E_{jx}\right|^2+\left|E_{jz}\right|^2\right)}{1+\varepsilon_j\left(\left|E_{jx}\right|^2+3\left|E_{jz}\right|^2\right)}\left(\frac{E_{jz}^2E_{jx}^*}{k_x}-\frac{\left|E_{jz}\right|^2E_{jx}}{k_x^*}\right),\\
\frac{dE_{jx}}{dz}=ik_xE_{jz}-\frac{ik_j^2}{k_x}\varepsilon_j\left(\left|E_{jx}\right|^2+\left|E_{jz}\right|^2\right)E_{jz},
\end{gather*}
\end{widetext}
where the asterisk denotes complex conjugation.

In the quasi-electrostatic approximation, we can neglect the last terms of the above equations.
These terms are proportional to $k_j^2$.
Then, after setting $A_j=E_{jz}/E_{jx}$ and $\tilde{\varepsilon}_j=\varepsilon_j\left|E_{jx}\right|^2$, we find the equations
\begin{gather}
\frac{dA_j}{dz}=-ik_x\left(1+A_j^2\right)-i\left(k_x+k_x^*\right)\frac{\tilde{\varepsilon}_j\left(A_j^2-\left|A_j\right|^2\right)}{1+\tilde{\varepsilon}_j\left(1+3\left|A_j\right|^2\right)},\\
\frac{d\tilde{\varepsilon}_j}{dz}=i\tilde{\varepsilon}_j\left(k_xA_j-k_x^*A^*_j\right).
\end{gather}
Next, we decompose $A_j$ and $k_x$ into their real and imaginary parts according to $A_j=R_j+iI_j$ and $k_x=k_x^r+ik_x^i$, and then separate the corresponding equations. Thus, we obtain the system
\begin{gather*}
\label{96}
\frac{dR_j}{d\tilde{z}}=\frac{4\tilde{\varepsilon}_jR_jI_j}{1+\tilde{\varepsilon}_j\left(1+3R^2_j+3I^2_j\right)}+2R_jI_j+\gamma\left(1+R^2_j-I^2_j\right),\\
\label{97}
\frac{dI_j}{d\tilde{z}}=\frac{4\tilde{\varepsilon}_jI^2_j}{1+\tilde{\varepsilon}_j\left(1+3R^2_j+3I^2_j\right)}+2\gamma R_jI_j-\left(1+R^2_j-I^2_j\right),\\
\label{98}
\frac{d\tilde{\varepsilon}_j}{d\tilde{z}}=-2\tilde{\varepsilon}_j\left(I_j+\gamma R_j\right),
\end{gather*}
where $\tilde{z}=zk_x^r$, $\gamma=k_x^i/k_x^r$. In the weakly dissipative regime, when $\left|\gamma\right|\ll 1$, we expand $R_j$ and $I_j$ as
\begin{gather*}
R_j\approx R^{(0)}_j+\gamma R^{(1)}_j,\\
I_j\approx I^{(0)}_j+\gamma I^{(1)}_j,\\
\tilde{\varepsilon}_j\approx \tilde{\varepsilon}^{(0)}_j+\gamma\tilde{\varepsilon}^{(1)}_j,
\end{gather*}
where the coefficients $R^{(\kappa)}_j$, $I^{(\kappa)}_j$ and $\tilde{\varepsilon}^{(\kappa)}_j$ do not depend on $\gamma$ $(\kappa=0, 1)$.
The ensuing equations for the zeroth-order variables $R^{(0)}_j$ and $I^{(0)}_j$ describe the energy distribution between $E_{jz}$ and $E_{jx}$ in the dissipationless limit (in which $\sigma_r\equiv 0$, or alternatively $k_x^i\equiv 0$). Specifically, the zeroth-order equations read
\begin{gather}
\frac{dR^{(0)}_j}{d\tilde{z}}=\frac{4\tilde{\varepsilon}^{(0)}_jR^{(0)}_jI^{(0)}_j}{1+\tilde{\varepsilon}^{(0)}_j\left(1+3\left(R^{(0)}_j\right)^2+3\left(I^{(0)}_j\right)^2\right)}+2R^{(0)}_jI^{(0)}_j,\notag\\
\frac{dI^{(0)}_j}{d\tilde{z}}=\frac{4\tilde{\varepsilon}^{(0)}_j\left(I^{(0)}_j\right)^2}{1+\tilde{\varepsilon}^{(0)}_j\left(1+3\left(R^{(0)}_j\right)^2+3\left(I^{(0)}_j\right)^2\right)}-1-\left(R^{(0)}_j\right)^2+\left(I^{(0)}_j\right)^2,\notag \\
\label{b10}
\frac{d\tilde{\varepsilon}^{(0)}_j}{d\tilde{z}}=-2\tilde{\varepsilon}^{(0)}_jI^{(0)}_j.
\end{gather}

To facilitate the treatment of this system, we view $R^{(0)}_j$ and $I^{(0)}_j$ as a function of $\tilde{\varepsilon}_j^{(0)}$. Accordingly, we solve the following equations:
\begin{widetext}
\begin{subequations}\label{eq:RI-eps}
\begin{gather}
\label{106}
\frac{dR^{(0)}_j}{d\tilde{\varepsilon}^{(0)}_j}=-\frac{2R^{(0)}_j}{1+\tilde{\varepsilon}^{(0)}_j\left(1+3\left(R^{(0)}_j\right)^2+3\left(I^{(0)}_j\right)^2\right)}-\frac{R^{(0)}_j}{\tilde{\varepsilon}^{(0)}_j},\\
\label{107}
\frac{dI^{(0)}_j}{d\tilde{\varepsilon}^{(0)}_j}=-\frac{2I^{(0)}_j}{1+\tilde{\varepsilon}^{(0)}_j\left(1+3\left(R^{(0)}_j\right)^2+3\left(I^{(0)}_j\right)^2\right)}+\frac{\left(R^{(0)}_j\right)^2-\left(I^{(0)}_j\right)^2+1}{2\tilde{\varepsilon}^{(0)}_jI^{(0)}_j}.
\end{gather}
\end{subequations}
\end{widetext}

 The initial conditions imposed on the variables of Eqs.~\eqref{eq:RI-eps} read
\begin{subequations}\label{eq:init-conds}
\begin{gather}
\label{cond1}
I^{(0)}_j\left|_{\tilde{\varepsilon}_j^{(0)}=0}=\left(-1\right)^{j+1},\right.\\
\label{cond2}
R^{(0)}_j\left|_{\tilde{\varepsilon}^{(0)}_j=0}=0. \right.
\end{gather}
\end{subequations}
Next, we add Eqs.~\eqref{eq:RI-eps} and multiply the result by the common denominator.
This manipulation yields the following expression:
\begin{widetext}
\begin{equation*}
\tilde{\varepsilon}^{(0)}_j\left(1+\tilde{\varepsilon}^{(0)}_j+3\tilde{\varepsilon}^{(0)}_j\left[\left(R^{(0)}_j\right)^2+\left(I^{(0)}_j\right)^2\right]\right)\frac{d\left[\left(R^{(0)}_j\right)^2+\left(I^{(0)}_j\right)^2\right]}{d\tilde{\varepsilon}_j^{(0)}}
+\left(2\tilde{\varepsilon}^{(0)}_j+1\right)\left[\left(R^{(0)}_j\right)^2+\left(I^{(0)}_j\right)^2\right]+3\tilde{\varepsilon}^{(0)}_j\left[\left(R^{(0)}_j\right)^2+\left(I^{(0)}_j\right)^2\right]^2
-1-\tilde{\varepsilon}_j^{(0)}=0.
\end{equation*}
\end{widetext}

A key point here is to recognize that the last expression can be written in the form
\begin{equation}
\label{a11}
\frac{dA(u,v)}{d\tilde{\varepsilon}^{(0)}_j}=0,
\end{equation}
where
\begin{equation*}
	u=\left(R^{(0)}_j\left(\tilde{\varepsilon}^{(0)}_j\right)\right)^2+\left(I^{(0)}_j\left(\tilde{\varepsilon}^{(0)}_j\right)\right)^2, \quad v=\tilde{\varepsilon}^{(0)}_j,
\end{equation*}
and
\begin{equation*}
A(u,v)=-v+uv-\frac{1}{2}v^2+uv^2+\frac{3}{2}u^2v^2.
\end{equation*}
Integrating Eq.~\eqref{a11}, we obtain
\begin{equation*}
\begin{split}
\frac{3}{2}\left(\tilde{\varepsilon}^{(0)}_j\right)^2\left[\left(R^{(0)}_j\right)^2+\left(I^{(0)}_j\right)^2\right]^2+\left(\tilde{\varepsilon}^{(0)}_j+\left(\tilde{\varepsilon}^{(0)}_j\right)^2\right)\\
\times\left[\left(R^{(0)}_j\right)^2+\left(I^{(0)}_j\right)^2\right]-\tilde{\varepsilon}^{(0)}_j-\frac{1}{2}\left(\tilde{\varepsilon}^{(0)}_j\right)^2=C
\end{split}
\end{equation*}
where $C$ is an integration constant.

Conditions~\eqref{eq:init-conds} imply that $C=0$ along with
\begin{equation*}
\left(R^{(0)}_j\right)^2+\left(I^{(0)}_j\right)^2=\frac{1}{3\tilde{\varepsilon}_j^{(0)}}\left(\sqrt{4\left(\tilde{\varepsilon}^{(0)}_j\right)^2+8\tilde{\varepsilon}_j^{(0)}+1}-\tilde{\varepsilon}_j^{(0)}-1\right).
\end{equation*}
By using the last relation and Eq.~\eqref{106}, we find that
\begin{equation*}
R^{(0)}_j=C_1\left(\tilde{\varepsilon}_1^{(0)}\right)^{-1}\left(\sqrt{4\left(\tilde{\varepsilon}_1^{(0)}\right)^2+8\tilde{\varepsilon}_1^{(0)}+1}+2\tilde{\varepsilon}_1^{(0)}+2\right)^{-1}
\end{equation*}
where $C_1$ is a constant. Then we apply Eq.~\eqref{cond2} to obtain $R^{(0)}_j=0$ and
\begin{equation}
\label{b21}
I^{(0)}_j=\left(-1\right)^{j+1}\sqrt{\frac{\sqrt{4\left(\tilde{\varepsilon}_j^{(0)}\right)^2+8\tilde{\varepsilon}_j^{(0)}+1}-\tilde{\varepsilon}_j^{(0)}-1}{3\tilde{\varepsilon}_j^{(0)}}}.
\end{equation}
This formula concludes our calculation of zeroth-order quantities $R^{(0)}_j$ and $I^{(0)}_j$.

We now turn our attention to $R^{(1)}_j$ and $I^{(1)}_j$ which account for the effect of small yet nonzero dissipation in the 2D material, assuming that $\sigma_r\ne 0$ and $\left|\sigma_r\right|\ll \left|\sigma_i\right|$. These $R^{(1)}_j$ and $I^{(1)}_j$ together with $\tilde{\varepsilon}_j^{(1)}$ satisfy the system of equations
\begin{gather}
\label{b22}
\frac{dR^{(1)}_j}{d\tilde{z}}=\frac{4\tilde{\varepsilon}_j^{(0)}R^{(1)}_jI^{(0)}_j}{1+\tilde{\varepsilon}_j^{(0)}+3\tilde{\varepsilon}_j^{(0)}\left(I^{(0)}_j\right)^2}+2R^{(1)}_jI^{(0)}_j+1-\left(I^{(0)}_j\right)^2,\\
\label{b24}
\frac{d\tilde{\varepsilon}_j^{(1)}}{d\tilde{z}}=-2\left(\tilde{\varepsilon}_j^{(0)}I_j^{(1)}+
\tilde{\varepsilon}_j^{(1)}I_j^{(0)}\right),
\end{gather}
\begin{widetext}
\begin{equation}
\label{b23}
\frac{dI^{(1)}_j}{d\tilde{z}}\left(1+\tilde{\varepsilon}_j^{(0)}+3\tilde{\varepsilon}_j^{(0)}\left(I^{(0)}_j\right)^2\right)^2=
2I^{(0)}_jI^{(1)}_j\left[1+6\tilde{\varepsilon}_j^{(0)}+5\left(\tilde{\varepsilon}_j^{(0)}\right)^2+6\tilde{\varepsilon}_j^{(0)}\left(I^{(0)}_j\right)^2+6\left(\tilde{\varepsilon}^{(0)}_j\right)^2\left(I^{(0)}_j\right)^2+9\left(\tilde{\varepsilon}_j^{(0)}\right)^2\left(I_j^{(0)}\right)^4\right]+4\tilde{\varepsilon}_j^{(1)}\left(I^{(0)}_j\right)^2.
\end{equation}

The combination of Eqs.~\eqref{b10}, \eqref{b21} and~\eqref{b22}  yields
\begin{equation}
\label{b25}
\frac{d\tilde{R}^{(1)}_j}{dI^{(0)}_j}=\frac{8\tilde{R}^{(1)}_jI^{(0)}_j}{\left(1+\left(I^{(0)}_j\right)^2\right)\left(3\left(I^{(0)}_j\right)^2-1\right)}+\frac{2\left(\left(I^{(0)}_j\right)^2-1\right)\left(3\left(I^{(0)}_j\right)^4-6\left(I^{(0)}_j\right)^2-1\right)}{\left(1+\left(I^{(0)}_j\right)^2\right)^2\left(3\left(I^{(0)}_j\right)^2-1\right)^2}.
\end{equation}
In the above, we define $\tilde{R}^{(1)}_j=\tilde{\varepsilon}^{(0)}_jR^{(1)}_j$.
Accordingly, from the solution of Eq.~\eqref{b25} we derive the formula
\begin{equation*}
R^{(1)}_j=\left(-1\right)^{j+1}\frac{\left(1-3\left(I^{(0)}_j\right)^2\right)^2}{\left(I^{(0)}_j\right)^2-1}\left[\frac{4+\pi}{16}-\frac{1}{4\sqrt{3}}\tanh^{-1}\frac{\sqrt{3}\left(I^{(0)}_j-1\right)}{3I^{(0)}_j-1}-\frac{1}{4}\tan^{-1} I^{(0)}_j\right]+\left(-1\right)^{j+1}\frac{I^{(0)}_j\left(1-2\left(I^{(0)}_j\right)^2\right)}{\left(I^{(0)}_j\right)^2-1}.
\end{equation*}

The solution of Eqs.~\eqref{b24} and~\eqref{b23} that reduces to the known solution of the linear problem as $\varepsilon_j\rightarrow 0$ is described by $I^{(1)}_j=0$ and $\tilde{\varepsilon}^{(1)}_j=0$.
Hence, the relation between the electric field components is found to be
\begin{equation*}
E_{jz}=\left(-1\right)^{j+1}E_{jx}\left[iF\left(\varepsilon_j\left|E_{jx}\right|^2\right)+\frac{k_x^i}{k_x^r}G\left(\varepsilon_j\left|E_{jx}\right|^2\right)\right],
\end{equation*}
where $F(a)$ and $G(a)$ are defined by Eqs.~\eqref{eq:FG-def}. 
\end{widetext}

\end{appendix}

%

\end{document}